\documentclass[]{aa} 
\usepackage{graphicx}
\usepackage{natbib} 
\usepackage{amssymb,amsmath}
\usepackage{float} 
\usepackage{caption, subcaption}
\usepackage{fixltx2e}
\usepackage{xspace} 
\usepackage{url} 
\usepackage[table]{xcolor}
\usepackage{longtable}
\usepackage{lscape}
\usepackage{morefloats}

\bibpunct{(}{)}{;}{a}{}{,} 

\def\xmm{XMM-Newton\ } 
\def\sas{XMM-SAS} 
\def\epic{ EPIC\ }
\def\lbqs{LBQS 2212-1759}

\begin{document}

	\title{
	Distant galaxy clusters in a deep \xmm field \\within the CFTHLS
	D4\thanks{This work made use of observations made with ESO Very Large
	Telescope at the La Silla Observatory under the programmes
	072.A-0706(A), 073.A-0737(A), 079.A-0369(C), and
	080.A-0659(A)}\thanks{\xmm, an ESA science mission with instruments and
	contributions directly funded by ESA Member States and
	NASA}\thanks{Tables \ref{tab:pntsrc} and \ref{tab:allspec} are only
	available in electronic form at the CDS via anonymous ftp to
	cdsarc.u-strasbg.fr (130.79.128.5) or via
	\texttt{http://cdsweb.u-strasbg.fr/cgi-bin/qcat?J/A+A/}}\thanks{
	Full-resolution	images are available only in the published version of
	the manuscript in A\&A.} 
	}


   \author{A. de Hoon\inst{1} 
   \and G. Lamer\inst{1} 
   \and A. Schwope\inst{1}
   \and M. M\"uhlegger\inst{2} 
   \and R. Fassbender\inst{2} 
   \and H. B\"ohringer\inst{2} 
   \and M. Lerchster\inst{2} 
   \and A. Nastasi\inst{2} 
   \and R. \v{S}uhada\inst{6} 
   \and M. Verdugo\inst{2,7}
   \and J.P. Dietrich\inst{6}
   \and F. Brimioulle\inst{6}
   \and P. Rosati\inst{2}
   \and D. Pierini\thanks{Freelance astronomer}
   \and J.S. Santos\inst{3}
   \and H. Quintana\inst{4}
   \and A. Rabitz\inst{1}
   \and A. Takey\inst{1,5}
   }
          
   \offprints{A. de Hoon}

   \institute{
   Leibniz-Institut f\"ur Astrophysik Potsdam (AIP), An der
   Sternwarte 16, D-14482 Potsdam, Germany\\ \email{arjen@aip.de} 
   \and Max-Planck-Institut f\"ur extraterrestische Physik, 
   Giessenbachstra\ss{e}, 85748 Garching, Germany
   \and European Space Astronomy Center (ESAC)/ESA, Madrid, Spain 
   \and Departamento de Astronom\'a y Astrof\'isica, Pontificia Universidad 
   Cat\'olica de Chile, Casilla 306, Santiago 22, Chile
   \and National Research Institute of Astronomy and Geophysics (NRIAG),
   Helwan, Cairo, Egypt 
   \and University Observatory Munich, Ludwig-Maximillians University Munich,
   Scheinerstr. 1, 81679 Munich, Germany 
   \and Institut f\"ur Astronomie, Universit\"at Wien, Universit\"atsring 1,
   1010, Wien 
   }

   \date{received 17 Sept. 2012; accepted 07 Dec. 2012}

   \abstract
	{}
	{The \xmm Distant Cluster Project (XDCP) aims at the identification
	of a well defined sample of X-ray selected clusters of galaxies at
	redshifts $z \geq 0.8$. As part of this project, we analyse the deep
	\xmm exposure covering one of the CFHTLS deep
	fields to quantify the cluster content. We validate the optical
	follow-up strategy as well as the X-ray selection function.}
	{We searched for extended X-ray sources in archival \xmm \epic
	observations.  Multi-band optical imaging was performed to select high
	redshift cluster candidates among the extended X-ray sources. Here we
	present a catalogue of the extended sources in one the deepest
	$\sim250$ ksec \xmm fields targetting \lbqs\ covering $\sim 0.2\;
	\Box^\circ$. The cluster identification is based on deep imaging with
	the ESO VLT and from the CFHT legacy survey, among others. The
	confirmation of cluster candidates is done by VLT/FORS2 multi-object
	spectroscopy. Photometric redshifts from the CFHTLS D4 were utilised to
	confirm the effectiveness of the X-ray cluster selection method. The
	survey sensitivity was computed with extensive Monte-Carlo
	simulations.}
	{At a flux limit of $S_{0.5-2.0\,\text{keV}}\sim
	2.5\cdot10^{-15}\;\text{erg s}^{-1}$ we achieve a completeness level
	higher than 50\% in an area of $\sim 0.13\,\Box^\circ$. We detect six
	galaxy clusters above this limit with optical counterparts, of which 5
	are new spectroscopic discoveries.  Two newly discovered X-ray luminous
	galaxy clusters are at $z\geq1.0$, another two at $z=0.41$, and one at
	$z=0.34$.  For the most distant X-ray selected cluster in this field at
	$z=1.45$, we find additional (active) member galaxies from both X-ray
	and spectroscopic data.  Additionally, we find evidence of large-scale
	structures at moderate redshifts of $z=0.41$ and $z=0.34$.}
	{The quest for distant clusters in archival \xmm data has led to 
	detection of six clusters in a single field, making \xmm an outstanding
	tool for cluster surveys. Three of these clusters are at $z\geq1$, which
	emphasises the valuable contribution of small, yet deep surveys to
	cosmology. Beta models are appropriate descriptions of the cluster
	surface brightness when performing cluster detection simulations to
	compute the X-ray selection function. The constructed $\log N-\log S$
	tends to favour a scenario where no evolution in the cluster X-ray
	luminosity function (XLF) takes place. 
	}

	 \authorrunning{A. de Hoon et al.} 
	 \titlerunning{Distant clusters in a deep XMM field}

	 \maketitle

\section{Introduction}

Clusters of galaxies are the largest gravitationally bound objects in the
Universe and are tracers of the cosmic structure. Since their formation and
evolution depends sensitively on the cosmological parameters, they are
strong cosmological probes. Identification of clusters from X-ray surveys
is currently the best method of constructing well defined samples of galaxy
clusters for cosmological studies.  The X-ray emission of the intracluster gas
depends mostly on the square of the gas density. Owing to the peaked gas density
distribution of clusters, the X-ray emission is more compact than the
distribution of cluster galaxies, hence less affected by projection
effects. For these reasons the X-ray luminosity is a good proxy of the total
cluster mass.

The identification of ROSAT X-ray sources has resulted in large samples of
local clusters \citep[e.g.][REFLEX]{reflex} and a limited number of clusters
at intermediate redshifts $z>0.4$ \citep[e.g.][]{vik98}. Only five distant
clusters beyond $z=1$ have been found among the ROSAT sources, most of them in
the ROSAT Deep Cluster Survey \citep[RDCS]{rdcs}.

Among the large number of extended sources serendipitously detected by
\xmm \citep{axel,xcs,ali} one can expect a significant fraction of
distant clusters of galaxies. 
To date, about $40$ clusters beyond redshift $z \gtrsim 1$ and about $10$
$z\gtrsim 1.4$ are known, of which IDCS J1426.5+3508 at $z=1.75$
\citep{stanford12}, ClG J0218.3-0510 at $z=1.62$
\citep{papovich,tanaka,pierre}, XMMU J0338.8+0021 at $z=1.49$ \citep{ale},
LH146 at $z=1.75$ \citep{henry}, XMMU J0044.0-2033 at $z=1.58$ \citep{joana},
and XMMU J1007.4+1237 at $z=1.56$ \citep{renedistant} are the most distant
ones. There have also been claims of even more distant clusters at $z\sim2$
\citep{andreon, gobat}; however, these sources lack either a clear X-ray
detection or spectroscopic confirmation.

We are conducting a focussed project on the identification of distant
($z\geq0.8$) clusters from serendipitously detected \xmm sources, the \xmm
Distant Cluster Project \citep[XDCP,][]{messenger,lamer,renexdcp}. An early
success was our discovery of the most distant cluster of that time, XMMU
2235.3-2557 at $z=1.39$ \citep{mullis}. To date the XDCP has been contributing
half of the known sample of distant galaxy clusters. The sample contains 22
X-ray bright sources in the redshift range $0.9<z<1.6$ alone \citep{renexdcp}.

In the current paper we report on the extended X-ray sources found in the field
of the broad absorption line (BAL) QSO \lbqs. This object has been repeatedly
targetted with \xmm, so that the field features one of the deepest \xmm \epic
(European Photon Imaging Camera) exposures taken at that point. This field also
received deep optical imaging coverage by the ESO Imaging Survey (EIS field
XMM-07, \cite{eis}, \cite{eis2}) and the Canada-France-Hawaii Telescope (CFHT)
legacy survey (LS) deep field four (D4).  Furthermore, we have performed
extensive spectroscopic follow-up with VLT/FORS2 to confirm cluster membership.
We supply a complete X-ray source list of the stacked X-ray images for point
sources as \emph{Online material} in Table \ref{tab:pntsrc}. However, in this
paper, we focus on the content of only extended sources (Table \ref{extended}).

The paper is structured as follows. In Section \ref{sec:xmmobs} we describe
the X-ray source detection and the X-ray spectral analysis of the brightest
extended sources. Section \ref{sec:optical} gives an overview of the optical
data available for the field and describes the galaxies over-densities in
photometric redshift space derived from them. The spectroscopic confirmation
of the cluster candidates is presented next in Section \ref{sec:spec}. Section
\ref{sec:id} presents the optical identifications of the extended X-ray
sources and the cluster redshifts. In the same section, the identification of
the individual objects is discussed. The cosmological interpretation of our
results is finally presented in Section \ref{sec:cosmo}.

We adopt a cosmology of H$_0=70\rm km\ s^{-1}\ Mpc^{-1}$,
$\Omega_\text{M}=0.3$, and $\Omega_\Lambda =0.7$ and magnitudes are in the AB
system.

\section{\xmm observations} 
\label{sec:xmmobs}

\subsection{X-ray data reduction and source detection} 
\label{sec:srcdetect}

\xmm has repeatedly targetted (yet never detected) the BAL QSO \object{\lbqs}
\citep{clavel}. The field received a total good exposure of $250$ ksec and is
thus one of the deepest XMM fields. We retrieved the \epic data sets from the
public XMM archive (XSA) and processed the data from observation data files
(ODFs) with the XMM Science Analysis Software ({\tt \sas\ v.
6.5})\footnote{\sas\ version {\it 6.5} was the prototypical release used to
construct the 2XMM catalogue \citep{watson}}.  Source detection runs with one
of the latest versions {\tt \sas\ v. 11} show no major changes with respect to
the source list used in this work. We therefore remain with the older version
for consistency and comparison with the 2XMM catalogue. 

To screen the data for periods of high particle background, we created light
curves of the high energy (7-15 keV) events for each of the three \epic
cameras. Periods where the count rate was more than 30\% higher than the median
rate were excluded from the later analysis. Table \ref{obssum} shows the
remaining exposure times of cleaned data for the individual pointing.

\begin{table}[t] 
  \centering
  \caption{\label{obssum}  Summary of \xmm observations.}
  \begin{tabular}{llrrr}\\ 
	\hline 
	ObsID      & Date    &	\multicolumn{3}{c}{good exposure time[s]}           \\ 
               &         & MOS1 & MOS2 & PN  \\  
	\hline 
	0106660101 & 2000-11-17 & 57229  & 57273  & 54829  \\
	0106660201 & 2000-11-18 & 52177  & 52073  & 38247  \\
	0106660401 & 2001-11-16 & 33350  & 33561  & -     \\
	0106660501 & 2001-11-17 & 8063   & 8091   & 5665  \\
	0106660601 & 2001-11-17 & 100790 & 101211 & 84178 \\
	\hline
	total      &            & 251609 & 252209 & 182919 \\
	\hline 
  \end{tabular} 
\end{table}

The event lists of each observation were transformed to a common astrometric
frame (using the {\tt\sas} task {\tt attcalc}) and images in the five standard
\xmm energy bands (\cite{watson}, \textit{band 1}: $0.2-0.5\,\text{ keV}$,
\textit{band 2}: $0.5-1.0\,\text{keV}$, \textit{band 3}: $1.0-2.0\,\text{
keV}$, \textit{band 4}: $2.0-4.5\,\text{keV}$, \textit{band 5}:
$4.5-12.0\,\text{keV}$) were binned for each camera and exposure. These bands
differ from the energy bands (with a focus on $0.35-2.4\,\text{keV}$) usually
used within the XDCP collaboration \citep{renexdcp}. 

For each of these images, an exposure map, background map, and detection mask
were created using the respective {\tt \sas} source detection tasks.
Subsequently, the science images, exposure maps, and background maps of the
individual exposures were added, creating one image for each of the three
cameras and five energy bands.

In the softest band ($0.2-0.5\,\text{keV}$) some of the \epic detectors show
spatially and temporally variable background features that can lead to
spurious detections of extended sources. On the other hand, the thermal
spectra of galaxy clusters only have a small fraction of their count rates in
the hardest band beyond $4.5\,\text{keV}$. Therefore we restricted the
detection of extended sources to the energy bands 2-4.

Source detection with the {\tt \sas} tasks {\tt eboxdetect} and {\tt emldetect}
was performed simultaneously on the nine images from three cameras and three
energy bands. The task {\tt eboxdetect} applies a sliding box detection
algorithm to the input images. The resulting list is passed to the task {\tt
emldetect}, which fits the calibration PSF, optionally convolved with an extent
model, to each input source. Extended sources were modelled using a King
profile of the form 

\begin{equation}
  f(x,y)=\left(1+\frac{(x-x_0)^2+(y-y_0)^2}{r_\text{c}^2}\right)^{-3/2}
  \label{eq:king}
\end{equation}

with core radii between $r_\text{c}=[4.0''-80.0'']$ as the model's extent
parameter space. $x-x_0$ and $y-y_0$ are pixel coordinates relative to the
central pixel of the object considered. The thresholds for the detection
likelihoods were set to $L=5.0$ for {\tt eboxdetect} and $L=6.0$ for {\tt
emldetect}, with $L=-\text{ln}(P_\text{ false})$. This corresponds to
probabilities of false detection of $P=6.7 \cdot 10^{-3}$ and $P=2.5 \cdot
10^{-3}$.

\onllongtabL{2}{%
\begin{landscape}
\begin{longtable}{ccccccccccc}
  \caption{\label{tab:pntsrc} Source list of all point sources from source
  detection on the stacked image generated by {\tt emldetect} as described
  in Section \ref{sec:srcdetect}. CTS stands for source photon counts and LIKE
  for detection likelihood.}\\
  \hline\hline
  id & XMMU J         &  RA       &   DEC     &    CTS    &CTS\_ERR &   RATE &   RATE\_ERR &   FLUX      &   FLUX\_ERR  & LIKE   \\
  &                &  (J2000)  &   (J2000) &           &         & $\text{s}^{-1}$ &
  $\text{s}^{-1}$  & $\text{erg s}^{-1}\,\text{cm}^{-2}$  &  $\text{erg  s}^{-1}\,\text{cm}^{-2}$  &    \\
  \endfirsthead
  \caption{continued.}\\
  \hline\hline
 id & XMMU J         &  RA       &   DEC     &    CTS    &CTS\_ERR &   RATE &   RATE\_ERR &   FLUX      &   FLUX\_ERR  & LIKE   \\
  &                &  (J2000)  &   (J2000) &           &         & $\text{s}^{-1}$ &
  $\text{s}^{-1}$  & $\text{erg s}^{-1}\,\text{cm}^{-2}$  &  $\text{erg  s}^{-1}\,\text{cm}^{-2}$  &    \\
 \hline
 \endhead
 \hline
 \endfoot
 \hline\hline
    1& 221515.1-173223& 333.812949& -17.539982&     7645.99&   97.58&   9.9986E-02&   1.3507E-03&   4.8201E-14&   8.5430E-16&14945.55\\
   2& 221550.2-175208& 333.959505& -17.869012&     6273.23&   89.97&   5.6898E-02&   8.4996E-04&   3.0027E-14&   5.6725E-16&10794.41\\
   3& 221523.5-174320& 333.848141& -17.722276&     6395.81&   92.54&   3.4883E-02&   5.1476E-04&   1.7462E-14&   3.4513E-16&10113.04\\
   4& 221623.3-174317& 334.097443& -17.721391&     5083.60&   82.01&   7.1080E-02&   1.2148E-03&   3.9088E-14&   7.9878E-16& 7965.78\\
   5& 221538.1-174632& 333.908757& -17.775786&     5224.04&   86.57&   2.9116E-02&   4.9236E-04&   1.5173E-14&   3.3547E-16& 7164.72\\
   6& 221603.0-174317& 334.012692& -17.721537&     3028.87&   66.49&   2.5590E-02&   5.8388E-04&   1.3759E-14&   3.8529E-16& 3734.52\\
   7& 221456.6-175052& 333.735954& -17.847859&     2744.57&   63.78&   2.7092E-02&   6.7580E-04&   1.4845E-14&   4.3134E-16& 3429.88\\
   8& 221623.4-174724& 334.097609& -17.790097&     3050.53&   66.61&   4.4460E-02&   1.0397E-03&   2.3940E-14&   6.6403E-16& 3669.26\\
\hline
\end{longtable}
\end{landscape}
}

\begin{figure}[htb] 
  \includegraphics[width=.5\textwidth]{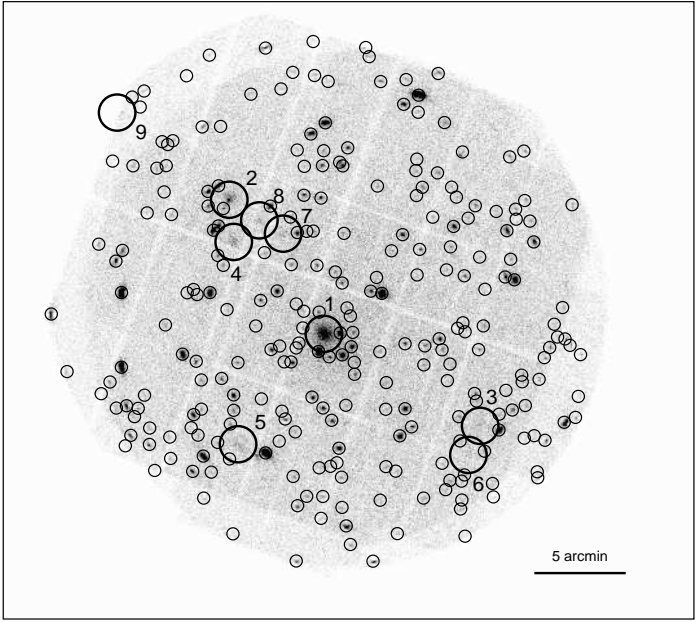}
  \caption{The stacked \epic image in the 0.2-2.0 keV band with detected point sources
  (small circles) and extended sources (large circles) labelled from 1 to 9.} 
  \label{fig:contour0}
\end{figure}

The {\tt emldetect} output source list contains 264 sources\footnote{The
complete source list of point source is found as \emph{Online material} in
Table \ref{tab:pntsrc}. The extended source, however, are listed in Table
\ref{extended}.}, of which nine are indicated to be extended. These objects are
numbered in Figure \ref{fig:contour0}. Table \ref{extended} lists the extended
sources with their basic X-ray parameters derived from \texttt{emldetect} and
sorted according to source counts. Spectral X-ray properties normalised to
physical scales are found in Table \ref{xrayiter}.

\begin{table*}[tb]
  \centering 
   \caption{\label{extended} Extended X-ray sources properties from source detection. CTS stands
   for photon counts, CR for count rate, S for flux, EXT\_LIKE is the extent
   likelihood, and DET\_LIKE the detection likelihood. The core radius is
   denoted by $r_\text{c}$.}
  \begin{tabular}{llrcccrr}
	\hline 
	\multicolumn{2}{c}{Source} & CTS & CR$_{0.5-4.5\,\text{keV}}$ &
	$S_{0.5-2.0\,\text{keV}}$ & $r_\text{c}$ & EXT\_LIKE & DET\_LIKE \\
	Seq. \# &  XMMU J..        &	 & $10^{-3}\,\text{s}^{-1}$ &
	$10^{-15}\,\text{erg cm}^{-2}\,\text{s}^{-1}$ & arcsec & & \\
	\hline 
	1 & 221536.8-174534 & 7982 & $42.3\pm1.00$ & $36.3\pm0.9 $ & $17.8\pm0.5$ & 696.6 & 2278.7  \\
	2 & 221558.6-173810 & 1807 & $17.3\pm0.70$ & $13.3\pm0.6 $ & $10.8\pm0.6$ & 154.0 & 580.4  \\
	3 & 221500.9-175038 & 1055 & $9.19\pm0.91$ & $8.65\pm0.73$ & $25.9\pm2.1$ & 42.2  & 80.5  \\
	4 & 221557.5-174029 & 909  & $8.79\pm0.65$ & $7.26\pm0.53$ & $14.9\pm1.1$ & 61.7  & 136.8  \\
	5 & 221556.6-175139 & 689  & $6.55\pm0.71$ & $5.69\pm0.58$ & $16.8\pm1.8$ & 26.3  & 61.6  \\
	6 & 221503.6-175215 & 485  & $4.75\pm0.74$ & $3.76\pm0.62$ & $19.4\pm3.0$ & 7.3   & 21.3  \\
	7 & 221546.2-174002 & 384  & $2.66\pm0.36$ & $1.88\pm0.29$ & $10.5\pm1.7$ & 6.6   & 24.8  \\
	8 & 221551.7-173918 & 384  & $3.32\pm0.41$ & $1.98\pm0.32$ & $ 9.3\pm1.5$ & 8.1   & 35.0  \\
	9 & 221624.3-173321 & 380  & $24.6\pm3.60$ & $29.9\pm4.3 $ & $20.1\pm3.9$ & 19.7  & 64.4  \\
	\hline 
   \end{tabular}
\end{table*}

\subsection{X-ray spectra and redshift estimates} 
\label{sec:xmmspec}

Thanks to the deep exposure in the field, several of the extended sources were
detected with more than $1\,000$ source counts (see Table \ref{extended}). We
therefore attempted to estimate the redshifts of the brightest cluster
candidates by fitting \epic spectra with plasma spectra templates, keeping
redshifts and temperatures as free parameters. With the same method we derived
metallicities for those clusters where it was feasible.

We extracted spectra (using {\tt \sas\ v.10.0}) for all the sources with $>400$
counts (i.e. sources ext\#1-\#6). Due to the low count rates ($0.01
-0.05\;\text{cts}\;\text{sec}^{-1}$) and the extended nature of the sources the
background fraction in the extracted spectra is relatively high. Therefore the
regions for source and background extraction were manually selected to avoid
systematic errors introduced by chip gaps, etc., and to avoid contamination by
unrelated X-ray sources. The data were rendered and fitted through {\tt XSPEC}
using C statistics. We applied single {\tt{\sc mekal}} models corrected for
galactic absorption ({\tt tbabs}).

The spectral fits enable us to get a handle on the clusters redshift, which
initially drove our follow-up programme. Some X-ray spectra enabled tight
constraints on the distance, as in the case of source ext\#1, which is the only
source alone the iron line has been detected. Others merely confirmed the
``high'' redshift $z\gtrsim 1$ nature of the cluster as displayed in Figures
\ref{fig:xray:ext1}-\ref{fig:xray:ext5}. An overview of spectral X-ray
properties is presented in Table \ref{xrayiter}. The individual results are
discussed both in Section \ref{sec:id} for each confirmed cluster individually
and in the appendix (Sect. \ref{sec:appendix}). The spectral properties in the
context of the proper flux determination are discussed in the next section.

\subsection{Flux determination}
\label{sec:flux}

To enable comparisons of our results with previous findings from other works
requires a flux measurement within a fixed radius. There, $S_{\Delta}$ is the
flux within $R_{\Delta}$, where $\Delta$ is defined as the average density of
the intracluster medium (ICM) relative to the critical density of the Universe
at a given redshift. The flux $S_{500}$ is commonly indicated, however, in most
cases not directly measurable, owing to the sensitivity of the instruments, the
background level, point source contamination and chip gaps.  The fluxes from
Table \ref{extended} are the results of the source detection, and they
correspond to fitted beta models integrated over a theoretically infinite
radius.

To obtain $S_\text{500}$, we apply an iterative approach. This method is
described in detail in \cite{suhada12} and \cite{ali}. To estimate $M_\text{
500}$ we apply the empirical scaling relation from \cite{vik09}. Since these
authors measure the flux in the energy band $0.5-2$ keV as well, no conversion
is required.  Appropriately, the evolution of the $L_\text{X}-M$ relation
deduced in \cite{vik09} is compatible with the findings from \cite{reichert11},
which both allow for a non-evolution scenario; i.e., the luminosity of clusters
does not change self-similarly with redshift. In the self-similar case, the
evolution factor is $E(z)^{-7/4}$. The physical parameters used in the
iterative approach are computed using the following equations:

\begin{align}
	M_\text{500} = 1.64\times 10^{-13} \, L_\text{500}^{0.62} \,
	E(z)^{-1.15}  \left(\frac{h}{0.72}\right)^{0.24} \label{eq:m500}\\
	R_\text{500} = \left( \frac{3 \, M_\text{500}}{4\pi \, 500 \, \rho
	(z)}\right) ^{\frac{1}{3}} \label{eq:r500}\\
	r_\text{c}   = 0.07 \, R_\text{500} \, T_\text{500}^{0.63} \label{eq:rc},
\end{align}
\noindent

where $\rho(z)$ is the critical density of the Universe, and $h$ the normalised
Hubble parameter. The evolution factor and $\rho(z)$ are defined as

\begin{align}
  E(z) = \sqrt{ \Omega_\text{m} \, (1+z)^3 + \Omega_\text{a}} \\
  \rho(z) = \frac{ 3 \, E(z)^2 \, H(z)^2 }{ 8 \, \pi \, G }
  \label{eq:evolution}
\end{align}

Given the luminosity from the spectral fit, we can make an estimation for
$M_\text{500}$ by applying equation \eqref{eq:m500} and setting $h$ to $0.70$.
The errors on the mass are the simple propagation of the luminosity
uncertainties. From this quantity we derive $R_\text{500}$ using equation
\eqref{eq:r500}. Since the cluster-specific core radius, which largely
determines the shape of the beta model, is only mildly realistically described
by the {\tt emldetect} output, we use the empirical relation, equation
\eqref{eq:rc}, from the cosmic evolution survey \citep[COSMOS,][]{cosmos} to
re-estimate $r_\text{c}$. The essential step in this procedure is to determine
what fraction of $R_\text{500}$ we have covered in the area from which the
spectrum has been extracted.

To do so, we synthesise beta models as in equation \eqref{eq:beta}. The beta
model is essentially a King profile from equation \eqref{eq:king}, and it
describes the surface brightness profile of a cluster, determined uniquely by
the core radius $r_\text{c}$, provided $\beta$ is known. By varying the radius
$R$, which is computed using equation \eqref{eq:r500}, we determine the
fraction of the surface brightness enclosed, which equals the luminosity
correction for a certain radius. 

\begin{table*}{btp}
  \centering
  \caption{\label{xrayiter} Extended X-ray sources. X-ray properties from
  spectral fitting and the iterative method (Sect. \ref{sec:flux}). Sources
  ext\#7-\#9 have been excluded, for they are not galaxy clusters. In case of
  non-symmetric errors, the largest deviation is presented. The first four
  columns display spectral measurements. $r_\text{ spec}$ is the radius in
  which the spectrum has been extracted. $R_\text{500}$, $L_\text{500}$ and
  $M_\text{500}$ are products of the iterative method.}
  \begin{tabular}{lclccllcc}
  \hline 
	   & $S_{0.5-2.0\, \text{keV}} $ & $L_{0.5-2.0\,\text{keV}}$ & $T$ & Abund.  &
	   $r_\text{spec}$ & $R_\text{500}$ & $L_\text{500}$ & $M_\text{500}$ \\
	   \#  & $10^{-15}\,\text{erg cm}^{-2}\,\text{s}^{-1}$ &
	   $10^{42}\,\text{erg s}^{-1}$ &
	   keV & $Z_{\sun}$ & arcsec & Mpc & $10^{42}\,\text{erg s}^{-1}$ & $10^{14}\,M_{\sun}$ \\
  \hline 
1 & $21.32\pm0.03$ & $11.7\pm0.02$  & $2.14\pm0.07$ &$0.34^{+0.06}_{-0.09}$ & 48 & 0.56 & $14.5\pm0.3$  & 0.79  \\
2 & $6.28\pm0.03$  & $54.0\pm1.50$  & $4.40\pm0.48$ & & 23 & 0.39 & $70.7\pm5.3$  & 1.07 \\
3 & $2.28\pm0.01$  & $0.86\pm0.002$ & $1.42\pm0.18$ & & 29 & 0.40 & $1.69\pm0.09$ & 0.17 \\
4 & $2.87\pm0.28$  & $13.4\pm0.13$  & $2.06\pm0.20$ & & 23 & 0.41 & $21.6\pm4.0$  & 0.57 \\
5 & $2.93\pm0.32$  & $25.1\pm0.27$  & $2.00\pm0.21$ & & 26 & 0.39 & $33.5\pm7.9$  & 0.71 \\
6 & $1.05\pm0.01$  & $0.59\pm0.07$  & $1.84\pm0.86$\tablefootmark{a} & & 30 & 0.34 & $0.92\pm0.2$  & 0.12 \\
  \hline
  \end{tabular}
 \tablefoot{\tablefoottext{a}{Tentative value, due to low signal-to-noise.}}
\end{table*}

When this quantity converges, i.e. $R=R_\text{500}$, after several iterations
(typically 2) we assume we have obtained the true $R_\text{500}$, and hence
$L_\text{500}$ proper. On average the spectral extraction radius has been $2.2$
times smaller than $R_\text{500}$. The resulting values are presented in Table
\ref{xrayiter} using spectral fluxes. 

\begin{equation} 
  S(R) = S_0 \left(1+ \left(\frac{R}{r_\text{c}}\right)^{2}
  \right)^{-3\beta+\frac{1}{2}}.
  \label{eq:beta} 
\end{equation}

We fix the temperature at the value from the spectral fit. We emphasise, that
all source detection chains in this work assume $\beta=\tfrac{2}{3}\eqsim
0.67$. The factor between $L_{0.5-2.0\,\text{keV}}$ and $L_\text{500}$ in Table
\ref{xrayiter} is applied to convert $S_{0.5-2.0\,\text{ keV}}$ into
$S_\text{500}$.

\section{Optical imaging} 
\label{sec:optical} 

\begin{figure}[htb]
    \centering
	\includegraphics[trim=0cm 0cm 1cm 0cm, clip=true, width=.5\textwidth]{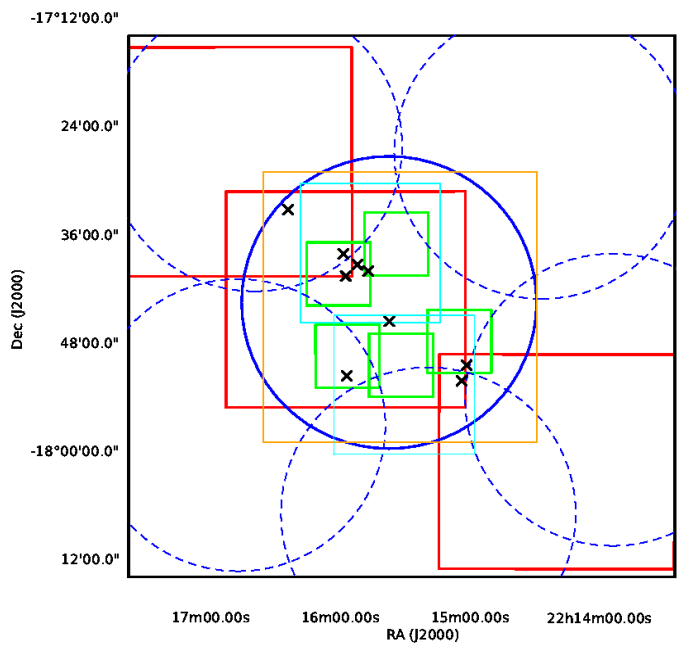}
	\caption{Sky coverage of imaging campaigns.{\it Black:} The optical
	$1\,\Box^\circ$ CFHTLS D4 field. {\it Orange:} The $BVRI$ sky
	projection as imaged by the WFI. {\it Red:} The NIR WIRcam
	supplementary coverage. {\it Cyan:} The NIR observations made at Calar
	Alto. {\it Blue:} \xmm pointings. The dashed circles represent the 2007
	study (P.I. K. Nandra) covering the entire D4 in X-rays with $\sim
	30-40\; \text{ks}$ exposures \citep{bielby}. {\it Green:} VLT-FORS2
	pointings used for initial detection. The crosses indicate the position
	of all clusters found. 
	}
  \label{fig:pointings}
\end{figure}

The field of \object{\lbqs} has been the target of various imaging programmes.
This section describes campaigns that have been launched as part of this study
and the archival data analysed in supplement. An overview of the sky coverage
is presented in Figure \ref{fig:pointings}. In this paper, only the area
enclosed by the central \xmm pointing (blue circle, solid line) is considered.

First, the field was selected for deep imaging observations in the $B,V,R$, and
$I$ bands with the Wide Field Imager (WFI) at the 2.2m ESO/MPIA telescope as
part of the ESO Imaging Survey \citep{eis}. The exposure time in the $R$-band
was augmented by a guest observer programme for the Bonn weak lensing survey
\citep[BLOX]{bolox}. In Oct. and Nov. 2006, a NIR campaign was run in the bands
$z$ and $H$ with the Omega 2000 camera at the Calar Alto 3.5m telescope. Both
ESO data sets have been used for the initial clusters identification. For all
final photometric analysis in this section, however, we rely solely on the data
from the CFHTLS, due to its supremacy. We describe the data reduction in
Section \ref{sec:cfht}. Section \ref{sec:photoz} describes the retrieval of
photometric redshifts, which we apply to find over-densities of galaxies in
Section \ref{sec:od}.

\subsection{CFHT data} 
\label{sec:cfht}

We make use of $ugriz$ archival data taken with the MegaPrime/MegaCam
instrument mounted at the prime focus at the 3.6m Canada-France-Hawaii
Telescope (CFHT). This instrument has a field of view of $1\times1$ deg$^2$ and a scale of 0.187 arcsec per pixel.  

We made use of data located in the deep field D4 of the CFHT Legacy Survey
(CFHTLS), which is part of the CFHT Supernova Legacy Survey (SNLS) and
observations are still ongoing.  We retrieved the data from the Elixir
system\footnote{http://www.cfht.hawaii.edu/Instruments/Elixir/home.html} in a
preprocessed form and further processed it as described in
\cite{erben09}\footnote{The full procedure of data reduction and calibration
will be described in Brimioulle et al. (2013, in prep.), see also
\cite{brimioulle08}.}. This reduction is independent of the official releases
by the CFHTLS collaboration \citep{ilbert,coupon}.  

The photometry was performed with  {\tt sextractor} \citep{bertin96} in dual
mode using the $i$-band image as a detection frame. We measured magnitudes in
apertures of 1.86 arcsec of diameter in seeing-matched images. The derived
magnitudes were used to derive colours, SED classification, and photometric
redshifts (see below). We also made use of deep near-infrared data ($JH$ \&
$Ks$ bands) with the WIRcam (also at CFHT). The data is part of the WIRcam
Deep Survey \citep[see][]{bielby} and covers a significant portion of most of
the deep XMM-Newton field. This data is important for obtaining precise
photometric redshifts at high redshift as important continuum features (e.g.
the $4\,000\,\text{\AA}$ break) are shifted to the near-infrared.

\subsection{Photometric redshifts}
\label{sec:photoz}

In this paper we use the multi-band imaging ({\it u$^*$g'r'i'z'JHKs}) and
photometric galaxy redshifts from the CFHTLS to identify cluster counterparts
of the extended X-ray sources and to measure their photometric redshifts.  We
used the code \texttt{PHOTO-z} \citep[][]{bender01} to estimate the photometric
redshifts, with a set of 31 SED templates \citep[see][for
details]{brimioulle08,mike}.

We used the results from the spectroscopic programme (Section \ref{sec:spec})
to calibrate our photometric redshift estimates by applying zero-point offsets
(typical values are $\sim 0.01..0.1$ mag). A total of $106$ spectra fulfil the
quality criteria (i.e. objects with reliable photometric \emph{and}
spectroscopic redshifts). The fraction of catastrophic outliers is defined as
$\eta=|z_\text{spec}-z_\text{phot}|/(1+z_\text{spec}) > 0.15$, which is $2.8
\%$ in our case.  The redshift accuracy is measured with the normalised median
absolute deviation, $\sigma = 1.48~ \times ~\text{median} (|\triangle z|/(1+z))
= 0.037$, where $\triangle z = z_\text{spec}-z_\text{phot} $ and $z$ is
spectroscopic. The value of $\sigma$ is identical to the standard deviation of
$|\triangle z|/(1+z)$.

\begin{figure}[t] 
  \centering
  \includegraphics[trim=.6cm .7cm .5cm .5cm, clip=true,width=.5\textwidth]{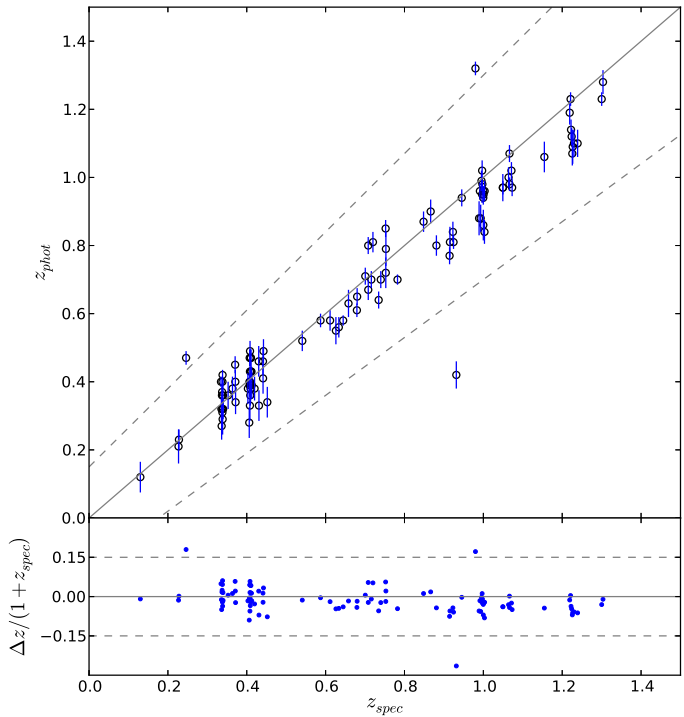}%
  \caption{Photometric redshifts, plotted against the spectroscopic ones. The
  vertical dashes are the photometric redshift errors. The
  dashed lines are for $z_{phot}=z_{spec}\pm 0.15 ~(1+z_{spec})$.}
  \label{FigPhotSpecComp} 
\end{figure}

\subsection{Galaxy over-densities}
\label{sec:od}

From the photometric redshift information we measured the projected galaxy
over-density in overlapping redshift intervals of $\Delta z=0.05$. The purpose
of this procedure was to qualitatively illustrate the completeness of the
X-ray selection technique as a method to identify clusters of galaxies.
Therefore we set a relatively high detection threshold in order to pick up
only the most significant density peaks.

\begin{figure*}[htb] 
  \centering
  \includegraphics[width=\textwidth]{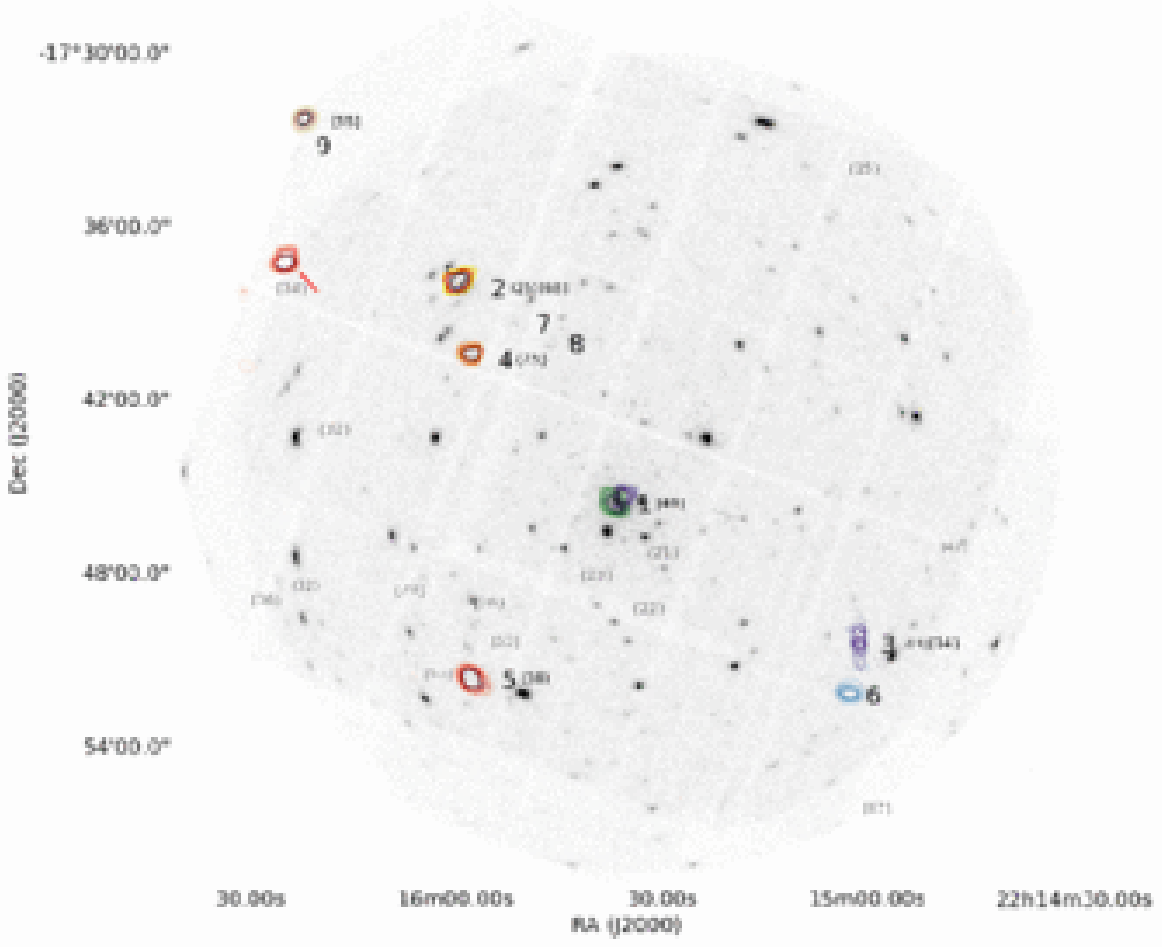}
  \caption{over-density contours of galaxies within the field of view of the
  \xmm pointing. The background image is the stacked image of all \xmm
  observations of LBQS2212 ($0.5-2.0$ keV). The extended X-ray sources from
  Figure \ref{fig:contour0} are indicated with their identifiers on the side
  ($1$ arcmin offset in R.A.). In smaller fonts the identifiers from other
  works are indicated either in black, if presented in this work as well, or in
  grey if not found by our method. Between round brackets the identifier from
  \cite{bielby} is indicated and square brackets refer to the catalogue from
  \cite{olsen}. Curly brackets present the clusters from \cite{adami10}. The
  red arrow points at an optical over-density, which has no clear counterpart
  in X-rays. The contours represent 5 Poissonian likelihood levels with
  $10<L<L_{max}$, colour-coded, or rather colour-mapped, by redshift interval:
  {\it purple} $z=0.3$ $L_{max}=15$, {\it blue} $z=0.35$ $L_{max}=20$, {\it
  green} $z=0.45$ $L_{max}=20$, {\it dark pink} $z=0.78$ $L_{max}=30$, {\it
  orange} $z=0.98$ $L_{max}=15$, {\it red} $z=1.18$ $L_{max}=20$, {\it heat}
  $z=1.38$ $L_{max}=20$, where $z$ refers to $z_{min}+\frac{1}{2}\Delta z$,
  i.e. the middle of the bin.} 
  \label{fig:zmaps} 
\end{figure*}

We computed the mean galaxy density in boxes of $400\; \text{kpc}$ a side,
which is enlarged to $500\; \text{kpc}$ for redshifts $z>0.9$, masking out
regions affected by bright stars. At redshifts beyond $z>0.9$ clusters are
thought to be increasingly less relaxed. Therefore, the need arises to search
for the associated galaxies in a larger physical volume. The values of $\Delta
z$, the box size and the critical redshift where the box size changes have been
tuned to maximise the detection of the already known clusters in this field
from this work. We consider only galaxies with a photometric redshift error
$\delta z < 0.1$. We compared the number of galaxies in the box to the assumed
Poisson distributed background, which is taken to be the mean galaxy count in
the field considered. The likelihood (equation \eqref{eq:like}) was computed
taking the negative natural logarithm of the incomplete gamma function $P(a,x)$
as defined in equation \eqref{eq:pois}. The incomplete gamma function is an
integral, which is applied to computing the likelihood that a given
distribution $x$ (i.e. the number of galaxies is the box) is exceeding
Poissonian noise $a$ \citep{cash}.

\begin{equation} 
  L=-\text{ln}\,P(a,x)
  \label{eq:like}
\end{equation} 
 with
\begin{equation} 
  P(a,x)=\frac{1}{\Gamma(a)} \, \int_0^x e^{-t} t^{a-1} dt, 
  \label{eq:pois}
\end{equation} 
\noindent

where $\Gamma(a)$ is the gamma function. This method of ``source finding'' is
similar to the \texttt{\sas} task \texttt{eboxdetect} \citep{watson} also
described in Section \ref{sec:srcdetect}. The resulting map is displayed in
Figure \ref{fig:zmaps} as contours on the stacked X-ray image. The redshift
intervals are chosen to overlap in order to prevent a search bias that would
occur when clusters fall exactly between two intervals. Since the redshift
error cut is set to $0.1$, some galaxy over-densities are expected to show up
in multiple adjacent redshift bins.  We have assumed a nominal redshift error
of $0.001$ for all galaxies, ensuring that each galaxy appears in one redshift
bin only. All confirmed X-ray selected clusters from Table \ref{tab:optical}
are retrieved at the appropriate redshifts. That is, all spectroscopically
confirmed clusters are found in a redshift interval around the nominal redshift
taking the spread in the photo-z accuracy into account (see Figure
\ref{FigPhotSpecComp}), of which typical values amount to about $5\%$.  

We define the photometric redshift $z_{phot}$ of the clusters as the
\emph{mean} of all galaxies meeting the above-mentioned selection criteria
(i.e. photometric accuracy and spatial distribution), having $\tfrac{\Delta
z_1}{1+z_{mid}} < 0.05$, where $\Delta z_1= \mid z^\ast-z_{mid}\mid$.
$z_{mid}$ is the middle of the redshift bin containing the largest number of
galaxies, whereas $z^\ast$ is the photometric redshift of each individual
galaxy. 
redshift.  The relative difference (in \%) in Table \ref{tab:optical} is
defined as $\tfrac{\Delta z_2}{1+z_{spec}}$, where $\Delta z_2=
z_{phot}-z_{spec}$. For the nearby clusters in our sample, $z_{phot}$ tends to
be slightly overestimated, while the high redshift ones the photometric
redshift is somewhat lower than the spectroscopic one. The redshift bin, in
which the galaxy signal is strongest (i.e. highest likelihood) is selected for
display in Figure \ref{fig:zmaps}. This quantity is identical neither to
$z_{mid}$ nor to $z_{phot}$ due to different treatments of background galaxies.
Nonetheless, they remain similar. 

The clusters are indicated with colour-coded contours reflecting the redshift
and the over-density likelihood. We display only significant\footnote{$L=10$
corresponds to a significance $>99.9\%$ or roughly $5\sigma$ in Gaussian
statistics.} over-densities having $L>10$, within the field of view of the \xmm
pointing. A selection of clusters found in previous publications
\citep{bielby,olsen,adami10} are also referred to in Figure \ref{fig:zmaps}.
The most distant \citep{stanford} cluster is detected by all groups except for
\cite{adami10}, whereas some other structures, i.e. system ext\#6 and the new,
possibly spurious structure (red arrow in Figure \ref{fig:zmaps}), are found
solely in this work. Source ext\#4 is also found by \cite{adami10} at
$z_{phot}=0.95$. Finally, we retrieve all X-ray selected clusters as optical
over-densities. Additionally, we find a significant over-density, that is
idicated by the arrow in Figure \ref{fig:zmaps}, which could not be associated
with an X-ray source at an off-axis angle $\sim 14$ arcmin. The optical data
suggest a detection due to an projection effect, since no conglomeration of
\emph{red} galaxies is evident.

\cite{olsen} have applied the Postman matched filter \citep[MF,][]{postman} to
search for optically selected clusters of galaxies in the CFHTLS deep fields.
In the overlap of field D4 with the \xmm FoV they found eight clusters with
redshifts ranging from $z=0.3$ to $z=1.1$. These structures are partially
retrieved by us as seen in Figure \ref{fig:zmaps}.  \cite{grove} have extended
the sample by including the $z'$-band. This has led to an enhanced detection of
distant clusters, which, however, are not discussed further in this work.  

In a similar fashion \cite{adami10} detect ten clusters within the common area
from galaxy density maps resulting from an adaptive kernel algorithm working on
slices in (photometric) redshift space.  The papers from \cite{olsen},
\cite{grove}, and \cite{adami10} are comparable to one another in various ways,
and for the same reasons different from our study. First of all, all base their
density map on a single band detection (for instance the $i'$ band), secondly,
use a similar detection algorithm for optical over-densities, and thirdly, none
perform an X-ray source detection, which automatically points them to a larger
quantity of less massive optical clusters. Last, the works addressed above are
restricted to the optical $u^*g'r'i'z'$ data and therefore are expected to
remain less sensitive to the more distant clusters in the field.  

\cite{gavazzi} use the $i'$-band images for weak lensing mass reconstruction.
In the D4 field they find one significant mass peak, which is outside the \xmm
FoV.  

A more direct camparison to our work is given by \cite{bielby}, who applied a
red sequence finder to their wavelet-detected extended X-ray sources in D4.
This team focussed on groups and clusters at high redshift $z\gtrsim 1.1$,
disgarding all sources at lower redshift. The number density of sources
($40\;\text{per}\; \Box^\circ$) they find is roughly equal to the results from
our search algorithm using \sas. These authors report on three distant X-ray
luminous groups within the area covered in this paper, of which two (see Figure
\ref{fig:zmaps}) have been retrieved by us as well both as X-ray and optical
over-density structures. Interestingly, \cite{bielby} make no reference to the
source ext\#4, described in Section \ref{sec:ext4}. One of their sources, ID-32
or WIRDXC J2216.4-1748, could not be confirmed by us. Likewise, we have been
able to spectroscopically refine the two photometric cluster redshifts found by
these authors.

\section{Optical spectroscopy}
\label{sec:spec}

The spectroscopic results used in this publication were obtained with
FORS2 mounted on the VLT through observing programme 079.A-0369(C) (Aug./Sept.
2007) and 080.A-0659(A) (Oct./Nov. 2007 and Jun. 2008). The total integration
time of roughly $19$ hours was used to target the nearby cluster candidates
ext\#1, ext\#3 and ext\#6 with three masks through grism \texttt{150I} and the
distant cluster candidates ext\#4 and ext\#5 with four masks through grism
\texttt{300I} grism. As spectroscopic targets we prioritised those objects
that lie close to the X-ray emission peak and/or have colours close to the
red sequence as determined from the ESO-VLT imaging. We also targetted some
X-ray point sources.

The science data was flat-fielded and wavelength-calibrated with the
appropriate calibration files supplied by the FORS2 database. The bias was
read from the chip overscan region. Flux calibration has been applied with
a single response curve derived from only one observation. No atmospheric
correction has been performed. The recorded spectra have a spectral coverage
from $6\,000-10\,000\;\text{\AA}$ ($\Re=660$) and $4\,000-10\,000\;\text{\AA}$
($\Re=260$) for the \texttt{300I} and \texttt{150I} respectively, where $\Re$
denotes the resolving power.

\begin{figure}[tb]
  \centering
  \includegraphics[width=.5\textwidth]{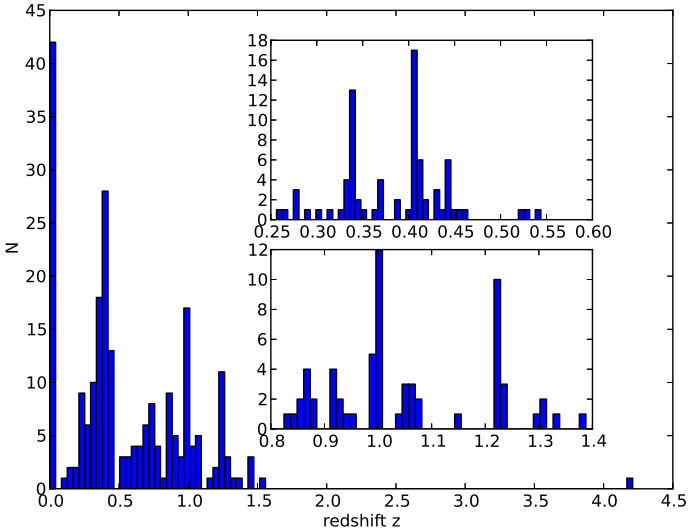}
  \caption{Histogram of all $231$ science grade spectra. The insets show a
  zoom-in on the low (top) and high (bottom) redshift range.}
  \label{fig:histz}
\end{figure}

From $289$ slits we were able to extract $231$ science-grade spectra, plus $19$
double recordings, having sufficient signal-to-noise to classify the spectra
and measure redshifts, of which six are QSOs and $42$ (late-type) stars. Their
redshift distribution is shown in Figure \ref{fig:histz}. A total of $66$
spectra are of cluster members, excluding ten double recordings. The remaining
$118$ spectra belong to unrelated field galaxies. The complete catalogue can be
viewed as \emph{Online material} in Table \ref{tab:allspec}. The list of
cluster members is presented in Table \ref{tab:spectra}. Galaxy cluster members
have been selected within an redshift interval $\Delta z\lesssim 0.015$ around
the mean redshift within a projected radius of $\lesssim 1\,\text{Mpc}$ around
the X-ray source.

\onllongtab{5}{%
\begin{longtable}{rcccllclccr}%
  \caption{\label{tab:allspec} Complete list of all spectra that have enough
  S/N, sorted according to the brightness in the $z'$ passband. The $z$ flag is
  defined as follows: (0) good (1) trustworthy (2) tentative. The template
  column indicates which template has been used to identify the object or
  determine its redshift. All stars listed are late types, primarily of the K
  and M classes.  The {\sf\sc id} is indicated for cluster members as they
  appear in Figures
  \ref{fig:sky:ext1},\ref{fig:sky:ext3},\ref{fig:sky:ext4},\ref{fig:sky:ext5},\ref{fig:sky:ext6}
  and in Table \ref{tab:spectra}.  The $z'$ and $H$ brightness are listed as
  {\tt AUTO} magnitudes. The magnitude flag is (0) when good and (1) when
  possibly affected by photometric effects.}\\
\hline\hline
\# & {\sf\sc id}& $\alpha$ & $\delta$  & $z$ & $\delta z$ & $z$ & template & $z'$ & $H$ & mag \\
   &            & (J2000) & (J2000) &     &            &flag &          &(mag) &(mag)&flag \\
\hline
\endfirsthead
\caption{continued.}\\
\hline\hline
\# & {\sf\sc id}& $\alpha$ & $\delta$  & $z$ & $\delta z$ & $z$ & template & $z'$ & $H$ & mag \\
   &            & (J2000) & (J2000) &     &            &flag &          &(mag) &(mag)&flag \\
\hline
\endhead
\hline
\endfoot
\hline
\hline
1 &           & 22:15:06.2 & -17:53:00.3 & 0 & $-$ & 0 &  star      & 15.72 & 15.16 & 1\\
2 &           & 22:15:47.5 & -17:39:20.7 & 0 & $-$ & 1 &  star      & 17.33 & 17.28 & 1\\
3 &           & 22:15:31.0 & -17:42:44.0 & 0 & $-$ & 0 &  star      & 17.40 & 16.92 & 0\\
4 &  ext\#3 B   & 22:15:01.4 & -17:50:35.9 & 0.3365 & 0.0001 & 0 &  passive   & 17.48 & 16.65 & 0\\
5 &  ext\#5 K   & 22:16:05.1 & -17:49:42.5 & 1.2231 & $<0.0001$ & 0 &  qso       & 17.66 & 17.56 & 0\\
6 &  ext\#6 A   & 22:15:03.6 & -17:52:10.1 & 0.4080 & $<0.0001$ & 0 &  passive   & 17.69 & 16.80 & 0\\
7 &           & 22:16:06.5 & -17:50:01.8 & 0.4597 & 0.0016 & 2 &  passive   & 17.77 & 18.02 & 0\\
8 &           & 22:15:50.6 & -17:43:54.0 & 0 & $-$ & 0 &  star      & 17.89 & 17.46 & 0\\
\hline
\end{longtable}
} 

Redshifts were obtained by correlating the spectra with the elliptical and
oldest starburst template from \cite{kin96}, both with the EZ software
\citep{gari10} for passive galaxies and the \textsc{\texttt{iraf fxcor}} task
for some galaxies with strong emission lines. 

Galaxy spectra were processed from $4\,500\,\text{\AA}$ up to
$10\,000\,\text{\AA}$, depending on spectral coverage and faulty regions.
Regions with strong sky absorption have been masked out. In most case this
range samples the following most prominent features: [OII] $\lambda\lambda
3726,3729$ emission lines, the CA H+K $\lambda\lambda 3934,3969$ absorption
lines, the $4\,000\,\text{\AA}$-break, the G-band around $\lambda\sim 4300$,
the hydrogen lines H$\delta$ $\lambda 4102$ and H$\beta$ $\lambda 4861$, and
[OIII] $\lambda\lambda 4959,5007$. Strong emission line galaxies were sampled
in a region around the most prominent emission lines. Only the two
mentioned templates were used to avoid systematic redshift shifts due to
resolution between templates.

No redshift errors are indicated in Table \ref{tab:spectra}, since the variance
in redshift resulting from using different templates (i.e. other templates than
the two mentioned above) is generally large ($\sim 0.001$) because of the
spectral resolution of the instrument, which is limited to $\delta z \sim
0.01$. The formal $1\,\sigma$ confidence intervals resulting from the $\chi^2$
template fit, however, are much smaller, typically $<0.001$. For emission line
galaxies the errors are smallest and are given by our spectral wavelength
calibration ($\sim0.1\,\text{\AA}$), which results in $\delta z\sim 0.00001$.
Line feature centroids can be determined with much higher precision than the
instrumental resolving power. In Table \ref{tab:allspec}, however, we merely
present the formal $1\sigma$ errors.

\begin{table}[htb] 
  \centering 
 \caption{\label{tab:spectra} Overview of spectroscopic redshifts for all
 confirmed cluster members with distances from the X-ray centre $\lesssim
 1\,\text{Mpc} / \lesssim 1.7\,\text{Mpc} / >2.5\,\text{Mpc}$ (white/grey/dark
 grey background).  The {\sf ID} corresponds to the identifiers in Figures
 \ref{fig:sky:ext1},\ref{fig:sky:ext3},\ref{fig:sky:ext4},\ref{fig:sky:ext5},\ref{fig:sky:ext6}
 in as far they appear within the cut-out. The entries are ordered according to
 their distance to the X-ray emission peak. The last row contains the mean for
 each column. Formal redshift errors ($1\,\sigma$) are typically $< 0.001$ and
 not listed here. A complete listing of all science-grade spectra can be found
 in Table \ref{tab:allspec}.} 
  \begin{tabular}{llllll}
	\hline 
	{\sf ID}    &  \multicolumn{5}{c}{redshift $z$} \\ 
	     & ext\#1 & ext\#3 & ext\#4 & ext\#5 &  ext\#6 \\ 
		 \hline 
  A & 0.407 & 0.337 & 0.999 & 1.237 & 0.408  \\
  B & 0.411 & 0.337 & 0.996 & 1.220 & 0.405  \\
  C & 0.409 & 0.335 & 1.001 & 1.222 & 0.408  \\
  D & 0.408 & 0.339 & 1.005 & 1.226 & 0.408  \\
  E & 0.406 & 0.337 & 0.992 & 1.223 & 0.413  \\
  F & 0.406 & 0.335 & 0.995 & 1.224 & 0.406  \\
  G & 0.410 & 0.338 & 1.003 & 1.229 & 0.409  \\
  H & 0.406 & 0.343 & 1.000 & 1.222 & 0.419  \\
  I & 0.408 & 0.337 & 1.001 & 1.228 & 0.410  \\
  J & 0.409 & 0.341 & 0.999 & \cellcolor[gray]{.8}1.225 & 0.403  \\
  K & 0.402 & 0.339 & 0.998 & \cellcolor[gray]{.8}1.223 & 0.409  \\
  L &   & 0.339 & 0.999 & \cellcolor[gray]{.8}1.240 & 0.407  \\
  M &   & 0.338 & \cellcolor[gray]{.8}1.001 &   & 0.410  \\
  N &   & 0.344 &   &   & 0.418  \\
  O &   & \cellcolor[gray]{.5}0.337 &   &   &    \\
  P &   & \cellcolor[gray]{.5}0.336 &   &   &    \\
  \hline
  mean & 0.408 & 0.338 & 0.999 & 1.227 & 0.410  \\
  \hline
 \end{tabular} 
\end{table}

\section{Identifications of \xmm cluster candidates} 
\label{sec:id}

\begin{figure}[tb]
  \centering
   \includegraphics[trim=.3cm 0cm 1cm 0cm,clip=true,width=.5\textwidth]{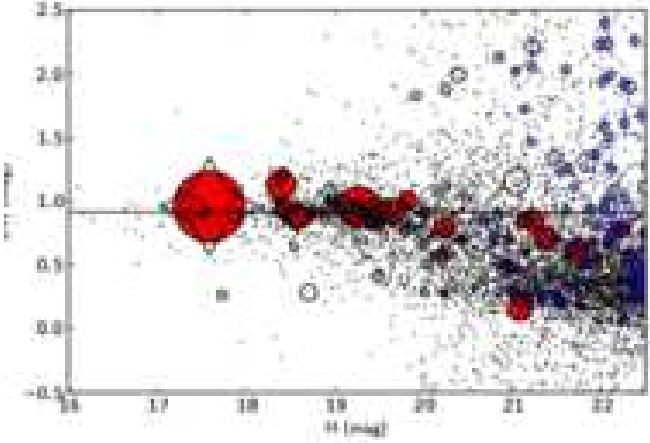}
   \caption{Colour-magnitude diagram for the extended X-ray
    source ext\#1.  CFTHLS D4 $z'$ and $H$-band images have been used. The small
    dots show \emph{all} galaxies within D4 having cluster-concordant photometric
    redshifts, i.e. within the area between the dashed lines in Fig.
    \ref{FigPhotSpecComp}.  Open circles show all galaxies with good photometry
    $<1$ Mpc from the X-ray peak, of which the filled symbols have
    cluster-concordant photometric redshifts. Vertical dashes indicate the error
    in the colour. Star symbols indicate spectroscopic members. All symbols
    (apart from the dots) have been scaled, inversely, to their physical distance
    from the X-ray emission peak. In each panel, the horizontal line indicates
    the expected colour of luminous, red sequence galaxies at the given
    spectroscopic redshift, as obtained from the simple stellar population models
    calibrated on data available in \cite{renexdcp}.}
   \label{fig:cmd:ext1}
  \end{figure}

\begin{figure}[tb]
  \centering
   \includegraphics[trim=.5cm 1cm 2.5cm 2cm,clip=true,width=.5\textwidth]{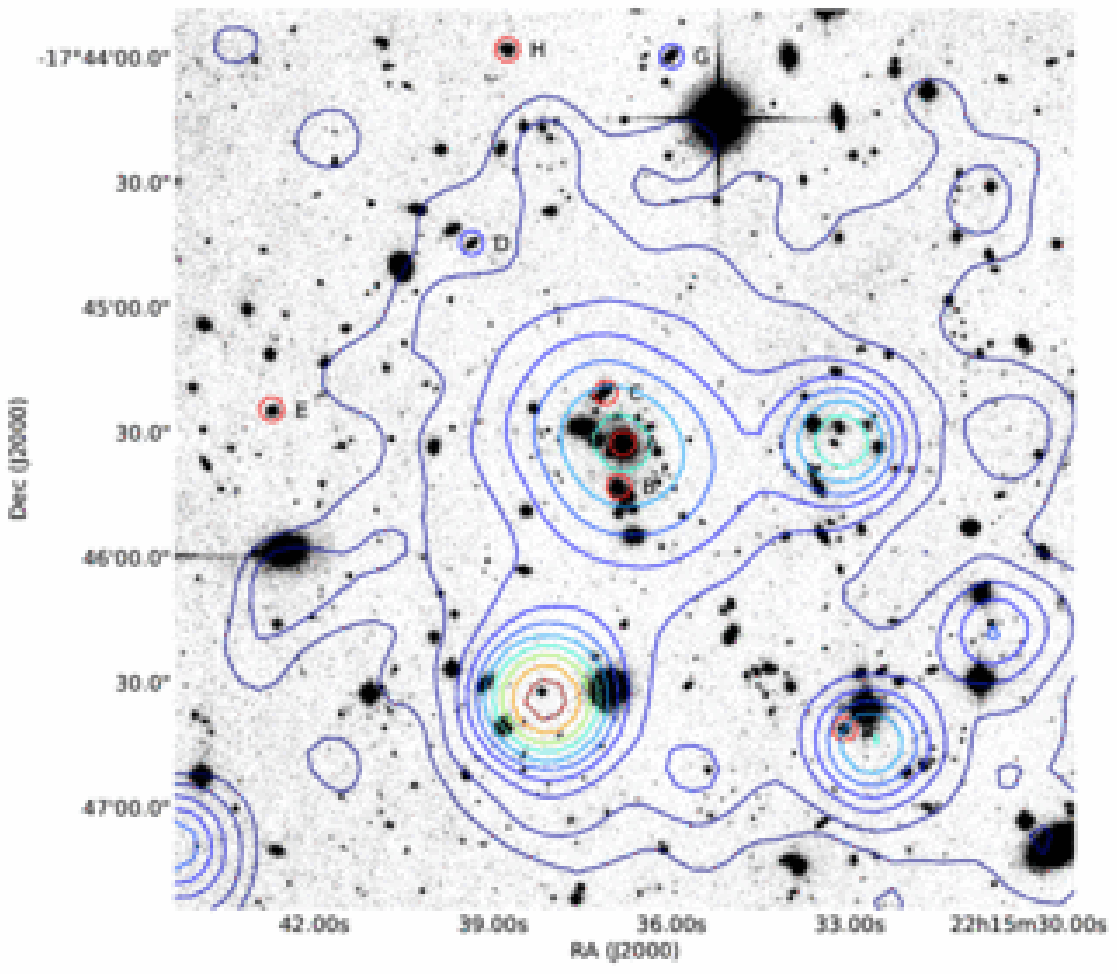}
   \caption{CFTHLS D4
    $z'$-band image centred on the extended X-ray sources in the field. The blue
    and red circles indicate the positions of the spectroscopically confirmed
    star forming and passive cluster members, respectively. The contours overlay
    represent linearly spaced photon count levels of the X-ray sources. North is
    up, east is to the right. The image is 1.2 arcmin across.}
    \label{fig:sky:ext1}
  \end{figure}

\begin{figure}[bt]
  \centering
   \includegraphics[trim=1cm 0cm 1cm 0cm,clip=true,width=.5\textwidth]{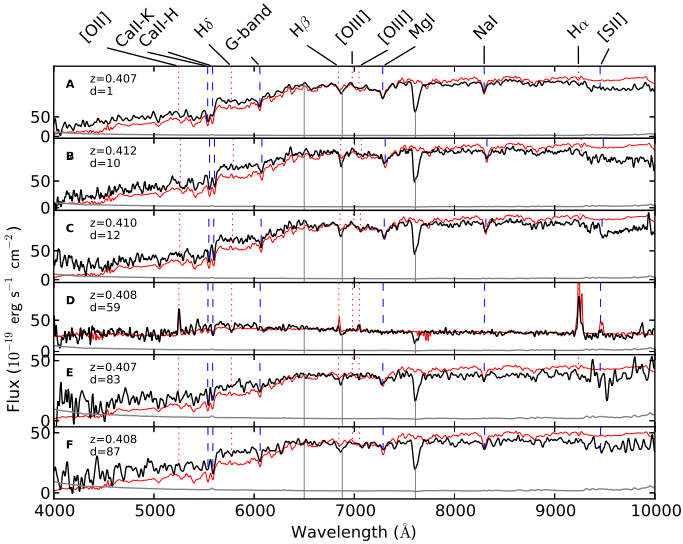} 
   \caption{Selected spectra for cluster ext\#1 in the observers' frame in black and
    their noise spectra in grey. Either an elliptical or emission template is
    overplotted in red. The positions of expected absorption and emission
    features have been indicated with blue dashed and red dotted lines,
    respectively. Vertical grey lines indicate regions affected by sky
    absorption. The spectra have been smoothed for clarity. The capital letters
    correspond to the entries in Table \ref{tab:spectra} and to the positions in
    Figure \ref{fig:sky:ext1}. The redshift {\sf z} and the distance {\sf d}
    (arcsec) to the formal peak of the X-ray source are indicated in the plot.}
    \label{fig:spec:ext1}
  \end{figure}

In the deep CFHT imaging six out of the nine extended X-ray sources can be
identified with clusters of galaxies. All mentioned redshifts are mean averaged
spectroscopically determined redshifts, unless stated otherwise.  Table
\ref{tab:optical} gives the likely identifications with their photometric
redshifts if available. Column ``redshift X-ray'' gives the 1$\sigma$ range of
the X-ray spectroscopic redshift estimates derived from $kT-z$ contours (see
Figures \ref{fig:xray:ext1}-\ref{fig:xray:ext5}). The cross-reference redshifts
listed in column ``redshifts publ.'' refer to the optically selected catalogues
from previous publication. In Figure \ref{fig:cmd:ext1}-\ref{fig:cmd:ext9} the
colour-magnitude diagrams $H_\text{ mag}$ vs. $(z'-H)_\text{mag}$ for all X-ray
positions are displayed, which are discussed qualitatively for each position
individually. The terms {\it red sequence} and the {\it expected colour} of a
cluster are used invariably, since we assume, as a zeroth-order approximation,
a zero gradient of $z'-H$ (mag) as a function of relative brightness. The
values of the empirically determined cluster colour as a function of redshift
are taken from \cite{renexdcp}. We chose to remain with a single colour, i.e.
$z'-H$, in order to illustrate its effectiveness thereof for the redshift's
range $0.3<z<1.5$ and to be able to relate our results to \cite{renexdcp}
(their Fig. 3). Yet, a single colour does not always optimally sample the
$4000\,\text{\AA}$ break, however, the chosen filter combination is a monotonic
function of the red sequence of passively evolving galaxies, hence sufficient
for our purpose.

\begin{table*}
  \centering
  \caption{\label{tab:optical} Optical identifications of extended sources.
  The column ``redshift publ.'' lists redshifts from already published works.
  The last column gives the difference, in percent, between the spectroscopic
  redshift and the photometric redshift of the BCG.}
  \begin{tabular}{lllcclllc}
	\hline 
	         & Source    &	Identification & \multicolumn{5}{c}{Redshift} & $\Delta z/(1+z)$ \\  
			 Seq. \#  &  XMMU J.. &                 & $z_\text{spec}$ & $z_\text{phot}$ & $z_\text{\textsc{bcg}}$ & publ. & X-ray & \% \\  
	\hline 
		1  & 221536.8-174534 & cluster & 0.41 & 0.46 & $0.47\pm0.05$ & $0.4^a$,$0.3^d$                        & 0.37-0.4 & $+3.7$  \\
		2  & 221558.6-173810 & cluster & -     & 1.34 & $1.37\pm0.06$ & $\underline{1.45}^b$,$1.37^e$, $1.2^a$ & 0.9-1.6  & $-4.5$  \\
		3  & 221500.9-175038 & cluster & 0.34 & 0.29 & $0.42\pm0.03$ & $0.3^a$,$0.38^e$                       & -        & $-3.4$
		\\
		4  & 221557.5-174029 & cluster & 1.00 & 0.90 & $0.98\pm0.04$ & $0.95^e$                               & 0.9-1.2  & $-5.0$
		\\
		5  & 221556.6-175139 & cluster & 1.23 & 1.12 & $1.23\pm0.04$ & $1.17^c$                               & $>0.9$   & $-4.8$   \\
		6  & 221503.6-175215 & cluster & 0.41 & 0.42 & $0.49\pm0.06$ & -                                      & -        & $+0.8$
		\\
		7                & 221546.2-174002                        & empty field  & -                        &      & - & - & - & -   \\
		8                & 221551.7-173918                        & empty field  & -                        &      & - & - & - & -   \\
		9                & 221624.3-173321                        & point source & -                        &      & - & - & - & -   \\
	\hline
  \end{tabular}
  \tablebib{$^a$\cite{olsen}; \underline{spectroscopic} redshift, $^b$\cite{stanford};
	$^c$\cite{bielby}; $^d$\cite{grove}; $^e$\cite{adami10}.} 
\end{table*}

\subsection{XMMU J221536.8-174534 (ext \#1) $z=0.41$} 
\label{sec:ext1}

Nearly on axis we find the brightest extended source at a redshift of
$z=0.408$. The position of brightest galaxy coincides with the X-ray emission
peak, as can be expected for relaxed clusters at relatively low redshifts. The
extended X-ray source has been described as alikely cluster of galaxies by
\cite{clavel}. This is confirmed by our optical imaging. The X-ray spectrum
(see Figure \ref{fig:xray:ext1}) is consistent with the photometric redshift
$z=0.4$ as found in the MF searches by \cite{olsen} and \cite{bolox}.  The CFHT
$z'$-band image with X-ray contours and spectroscopic members is shown in
Figure \ref{fig:sky:ext1}.
The cluster has a well-evolved red sequence as seen in Figure
\ref{fig:cmd:ext1} consistent with the empirically expected $z'-H$ colour. 
Six of its 11 spectra are presented in Figure \ref{fig:spec:ext1}. 

In Figure \ref{fig:zmaps} system ext\#1 is associated with two over-density
contours: the purple ones at $z\sim0.3$ and the greens ones at $z\sim0.45$. It
is possible, therefore, that the X-ray emission is a superposition of two
clusters at different redshifts. From the X-ray redshift-temperature contours
and from the positioning of galaxies with respect to the X-ray emission peak,
however, we are able to conclude that the X-ray emission originates in the more
distant redshift of the two. The structure at $z\sim 0.3$ was detected only as
galaxy over-density and is not regarded as contributing to the X-ray emission.
Additionally, we retrieved two spectroscopic redshifts at $z\sim 0.34$ at about
$50$ arcsec from the peak of source ext\#1, which are listed as object O and P
in the ext\#3 column in Table \ref{tab:spectra}. 

\subsection{XMMU J221558.6-173810 (ext \#2) $z=1.45$}
\label{sec:ext2}

\begin{figure}[tbh]
  \centering
  \includegraphics[trim=.3cm 0cm 1cm 0cm,clip=true,width=.5\textwidth]{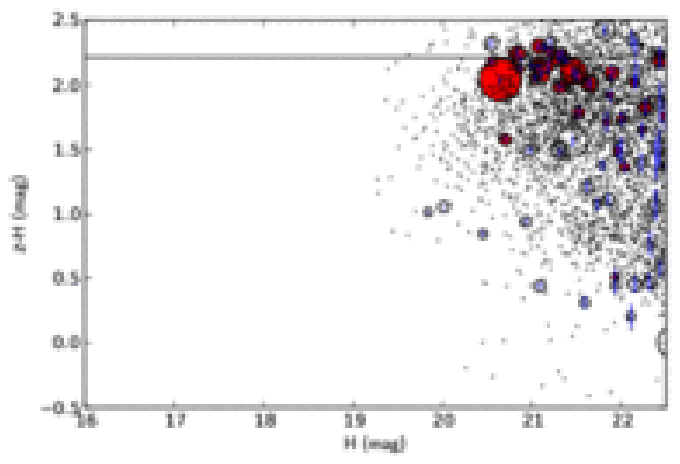}
  \caption{As in Figure \ref{fig:cmd:ext1} for source ext\#2 at $z=1.45$.}
  \label{fig:cmd:ext2}
\end{figure}

\begin{figure}[tbh]
  \centering
  \includegraphics[width=.5\textwidth, trim=.5cm 1cm 2.5cm 2cm, clip=true]{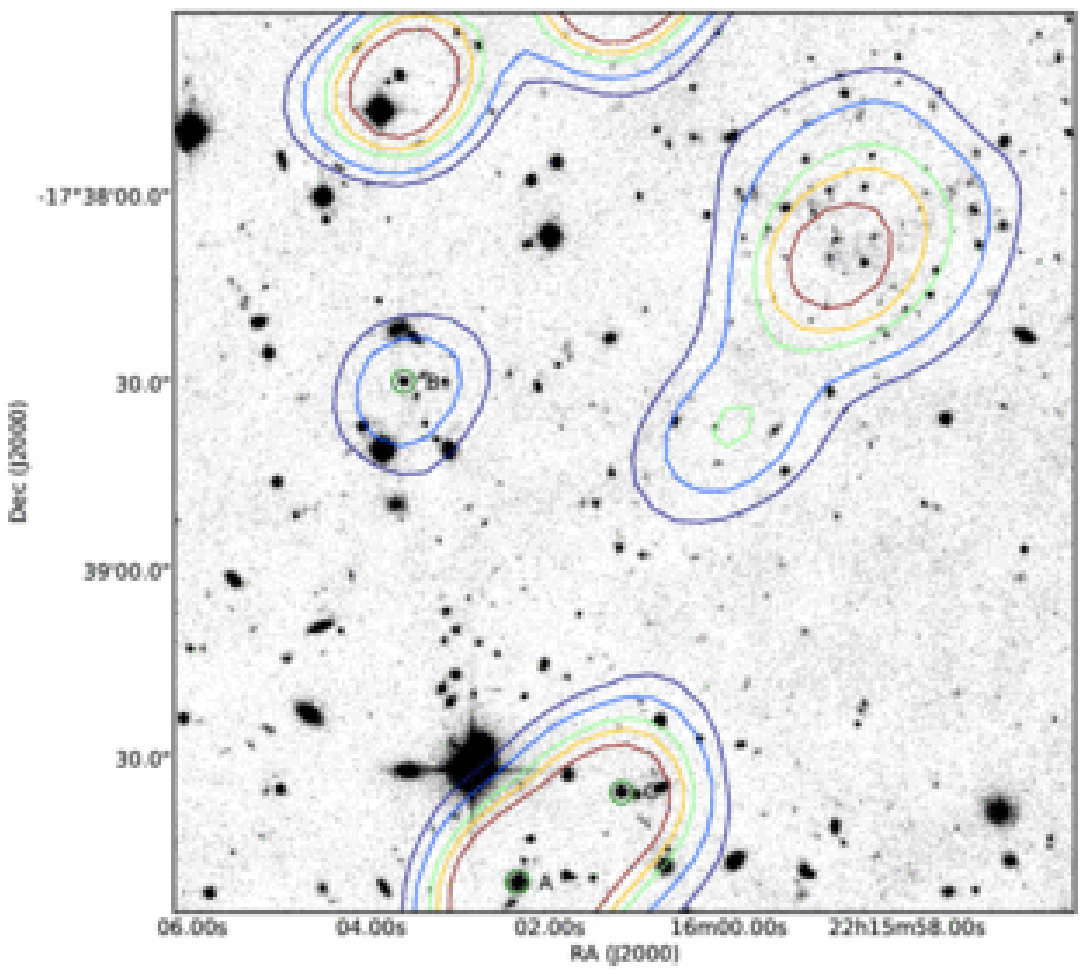}
  \caption{As in Fig. \ref{fig:sky:ext1} for source ext\#2. The cluster is NE
  of the image centre. No new spectra have been obtained for this publication,
  apart from the three labelled X-ray point sources, whose optical spectra are
  shown in Figure \ref{fig:spec:quasars}.}
  \label{fig:sky:ext2}
\end{figure}

The most distant galaxy cluster in this field has already been reported by
\cite{stanford}. It has been the topic of several dedicated studies since then,
e.g. studying star formation at high redshift \citep{hilton10, hayashi11}. This
is one of the most distant known X-ray selected cluster of galaxies with
unambiguously strong X-ray emission, albeit with a confirmed contamination from
point sources \citep{hilton09}. In addition to earlier publications, we find
several galaxies with photometric redshifts compatible with $z \sim 1.45$. The
over-density contours in Figure \ref{fig:zmaps} for this cluster are displayed
with the \emph{heat} colour scheme. An CFHT $z'$-band image is shown in Figure
\ref{fig:sky:ext2} with X-ray contour overlays and the positions of three new
quasars indicated, of which two reside at the cluster redshift.
The colour-magnitude diagram in Figure \ref{fig:cmd:ext2} shows that a red
sequence is already present at this high redshift, following the empirically
expected colour.
We have not obtained additional galaxy spectra in our observation runs.

\begin{figure}[tbh] 
  \begin{center}
   \includegraphics[trim=1cm 0cm 1cm 0cm, clip=true, width=.5\textwidth]{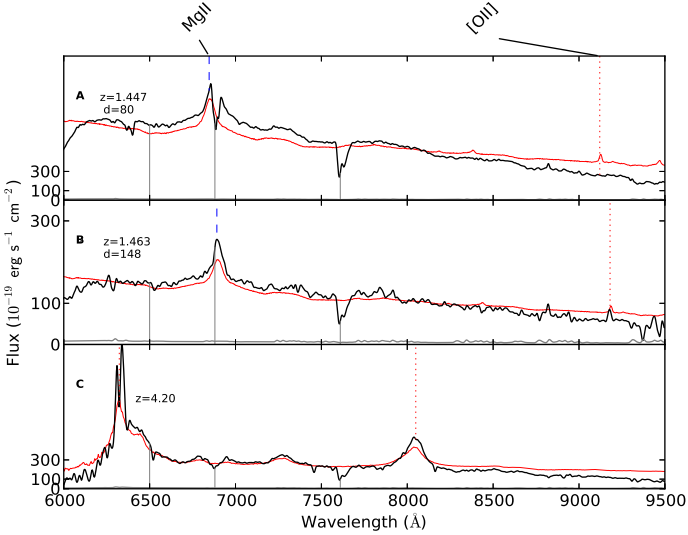} 
  \end{center} 
  \caption{Spectra (black) of the quasars discovered in the vicinity of
  cluster ext\#2. In red a quasar template is overplotted. Objects {\sc a} and
  {\sc b} can be regarded as cluster members or rather active galaxies in the
  cluster outskirts. Object {\sc c} is an unrelated distant X-ray luminous
  quasar serendipitously found in this study \citep{georg_poster}. The labels
  correspond to the objects identified in Figure \ref{fig:sky:ext2}. Spectral
  features, etc., are coded as in Figure \ref{fig:spec:ext1}. In the case of
  {\sc a} the emission lines are Ly$\alpha$ and CIV.} 
  \label{fig:spec:quasars} 
\end{figure}

However, we serendipitously identified two quasars at the nominal cluster
redshift on the very outskirts of the cluster (see Figure \ref{fig:sky:ext2}).
The spectra are presented in Figure \ref{fig:spec:quasars} (identifiers A and
B). For future spectroscopic X-ray analysis, these X-ray sources do not need to
be included as power laws contributing to the X-ray flux since they lay beyond
$1$ arcmin, i.e. farther than the spectral extracting radius. In a joint {\it
Chandra}-\xmm analysis, \cite{hilton10} report a temperature of $4.1$ kT with
the inclusion of three power laws to account for point source contributions. In
contrast, our analysis, as stated before, assumes the emission to result from
only a single hot plasma. The extraction region used by \cite{hilton10} is only
mildly (7 arcsec) larger than the one we used. 

\subsection{XMMU J221500.9-175038 (ext \#3) $z=0.34$} 
\label{sec:ext3}

\begin{figure}[bth]
  \centering
  \includegraphics[trim=.3cm 0cm 1cm 0cm,clip=true, width=.5\textwidth]{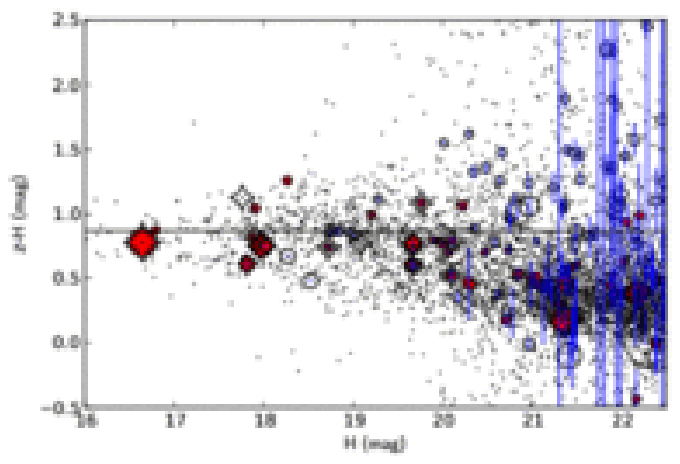}
  \caption{As in Figure \ref{fig:cmd:ext1} for source ext\#3 at $z=0.34$}
  \label{fig:cmd:ext3}
\end{figure}
  
\begin{figure}[tbh]
  \centering
  \includegraphics[width=.5\textwidth, trim=0.5cm 1cm 2.5cm 2cm, clip=true]{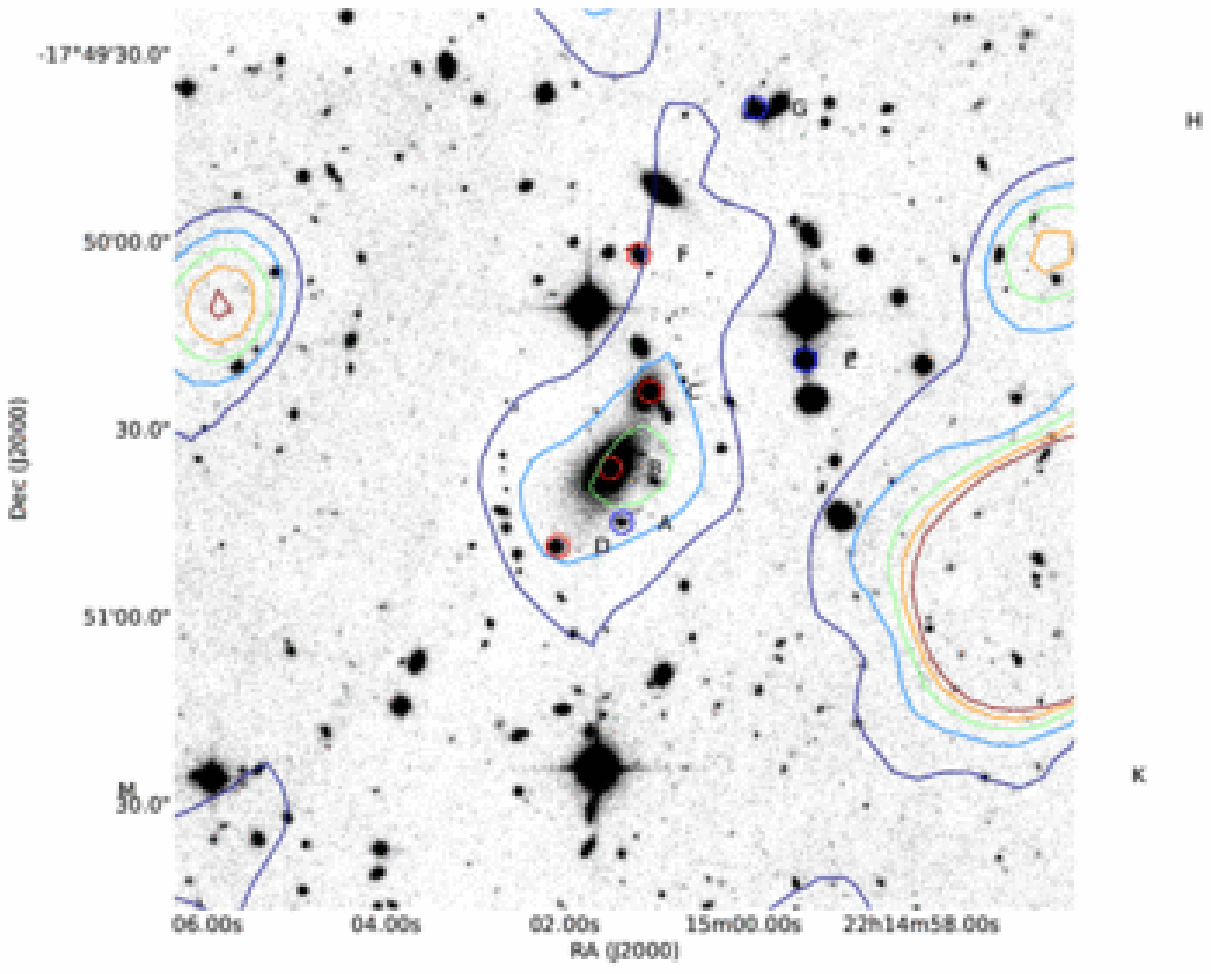}
  \caption{As in Fig. \ref{fig:sky:ext1} for source ext\#3. The associated
  optical spectra are found in Figure \ref{fig:spec:ext3}.}
  \label{fig:sky:ext3}

\end{figure}

\begin{figure}[bth]
  \centering
	\includegraphics[trim=1cm 0cm 1cm 0cm, clip=true, width=.5\textwidth]{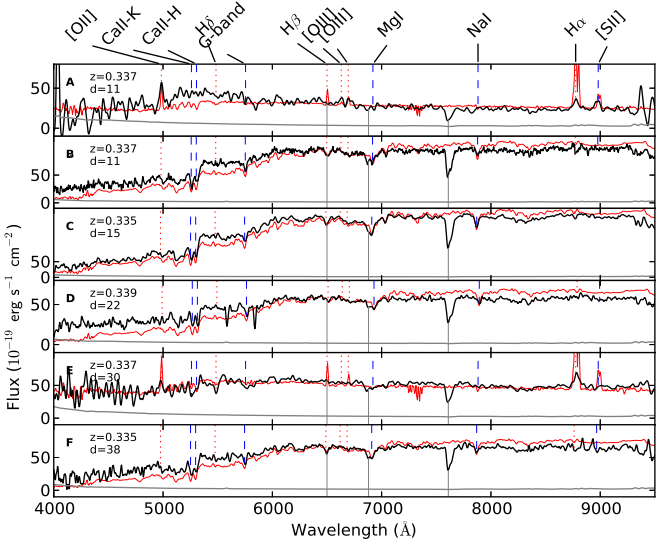}
  \caption{Selected spectra for cluster ext\#3 in the observers' frame. See
  Fig. \ref{fig:spec:ext1} for explanation. The positions are seen in
  Figure \ref{fig:sky:ext3}.} 
  \label{fig:spec:ext3} 
\end{figure}

At a mean redshift $z=0.338$ cluster ext\#3 is the closest galaxy cluster in the
field. This cluster was also detected in the MF searches by \cite{olsen}
($z\sim0.3$) and \cite{bolox}. We show the CFHT $z'$-band image with X-ray
emission contours and labelled spectroscopic members in Figure
\ref{fig:sky:ext3}. 
The red sequence is easily visible in Figure \ref{fig:cmd:ext3} at the colour
expected for its redshift.
Its selected $14$ spectroscopic members are within $\mid z-z_\text{cluster}\mid \lesssim 0.006$. Figure \ref{fig:spec:ext3} shows six of the confirmed members.  

In the same redshift bin as cluster ext\#3 (green contours in Figure
\ref{fig:zmaps}) we find an over-density of galaxies virtually at the same
projected position on the sky as the central cluster ext\#1. One interpretation
is to consider these over-densities to be a sign of knots in the cosmic web
$\sim 3$ Mpc apart; that is, the two galaxy over-densities are part of the same
large-scale structure. There is no X-ray detection from a group of galaxies at
$z=0.338$ at the coordinates of cluster ext\#1. Namely, the X-ray flux is
compatible with only coming from source ext\#1 as discussed in Section
\ref{sec:ext1}.

\subsection{XMMU J221557.5-174029 (ext \#4) $z=1.00$} 
\label{sec:ext4}

\begin{figure}[tbh]
  \centering
  \includegraphics[trim=.3cm 0cm 0cm 0cm,clip=true,width=.5\textwidth]{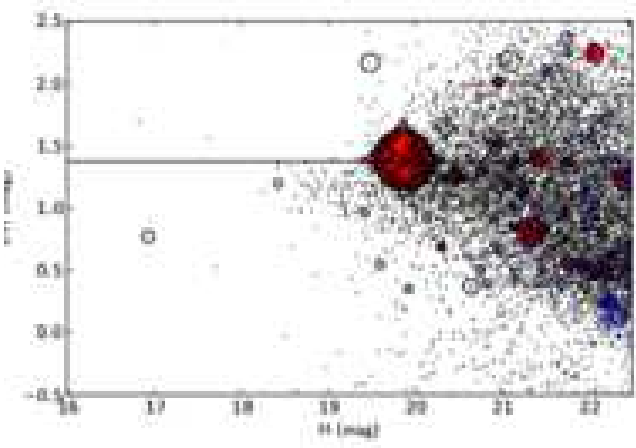}
  \caption{As in Figure \ref{fig:cmd:ext1} for cluster ext\#4 at $z=1.00$.}
  \label{fig:cmd:ext4}
\end{figure}

\begin{figure}[tbh]
  \centering
  \includegraphics[width=.5\textwidth, trim=0.5cm 1cm 2.5cm 2cm, clip=true]{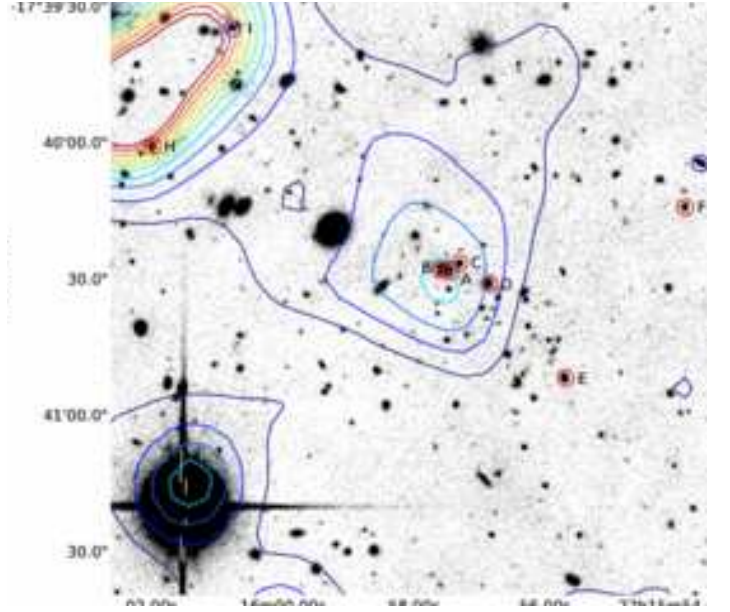}
  \caption{As in Fig. \ref{fig:sky:ext1} for source ext\#4. The associated
  optical spectra are found in Figure \ref{fig:spec:ext4}.}
  \label{fig:sky:ext4}
\end{figure}

On the very edge of the redshift unity barrier we find a new cluster with $13$
members. This cluster has already been referred to by \cite{adami10} as a
photometric cluster candidate; however, it appears in no other catalogue. Two
members are at the very centre of the X-ray emission, and both can be
considered BCG, since their fluxes are similar.

\begin{figure}[htb] 
  \begin{center}
   \includegraphics[trim=1cm 0cm 1cm 0cm, clip=true, width=.5\textwidth]{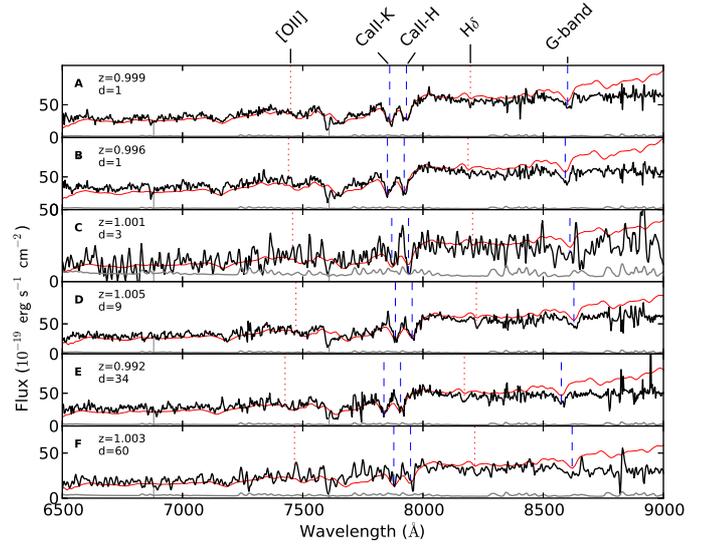} 
  \end{center}
  \caption{Selected spectra for cluster ext\#4 in the observers' frame. Colours
  and labels are as in
  Figure \ref{fig:spec:ext1}. Labels refer to identified objects in Figure
  \ref{fig:sky:ext4}. The spectra have been smoothed.} 
  \label{fig:spec:ext4} 
\end{figure}

The red sequence is visible in Figure \ref{fig:cmd:ext4} without deviations
from the expected $z'-H$ colour. Striking, however is the absence of a clear
magnitude gap between the BCGs and the other galaxies, which could indicate a
recently formed system.
The spectra, a selection of six from 13, are presented in Figure
\ref{fig:spec:ext4}. The two BCGs have spectroscopic redshifts $\Delta z =
0.003$, which is well beyond their formal ($1\sigma$) errors.
The cluster redshift of $z\sim 1.0$ is consistent with the estimate from the
X-ray spectrum. An image of the cluster is shown in Figure \ref{fig:sky:ext4}
with X-ray contours and confirmed members. The apparent point-like morphology of
the X-ray emission is purely due to chip gaps masking out the cluster edges. 

\subsection{XMMU 221556.6-175139 (ext \#5) $z=1.23$} 
\label{sec:ext5}

\begin{figure}[bth]
  \centering
  \includegraphics[trim=.3cm 0cm 1cm 0cm,clip=true,width=.5\textwidth]{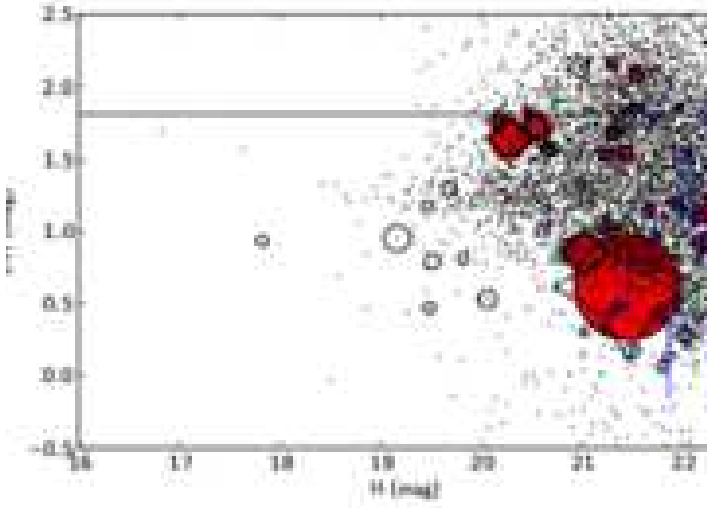}
  \caption{As in Figure \ref{fig:cmd:ext1} for cluster ext\#5 at $z=1.23$.}
  \label{fig:cmd:ext5}
\end{figure}

\begin{figure}[b] 
  \centering 
  \includegraphics[width=.5\textwidth, trim=0.5cm 1cm 2.5cm 2cm, clip=true]{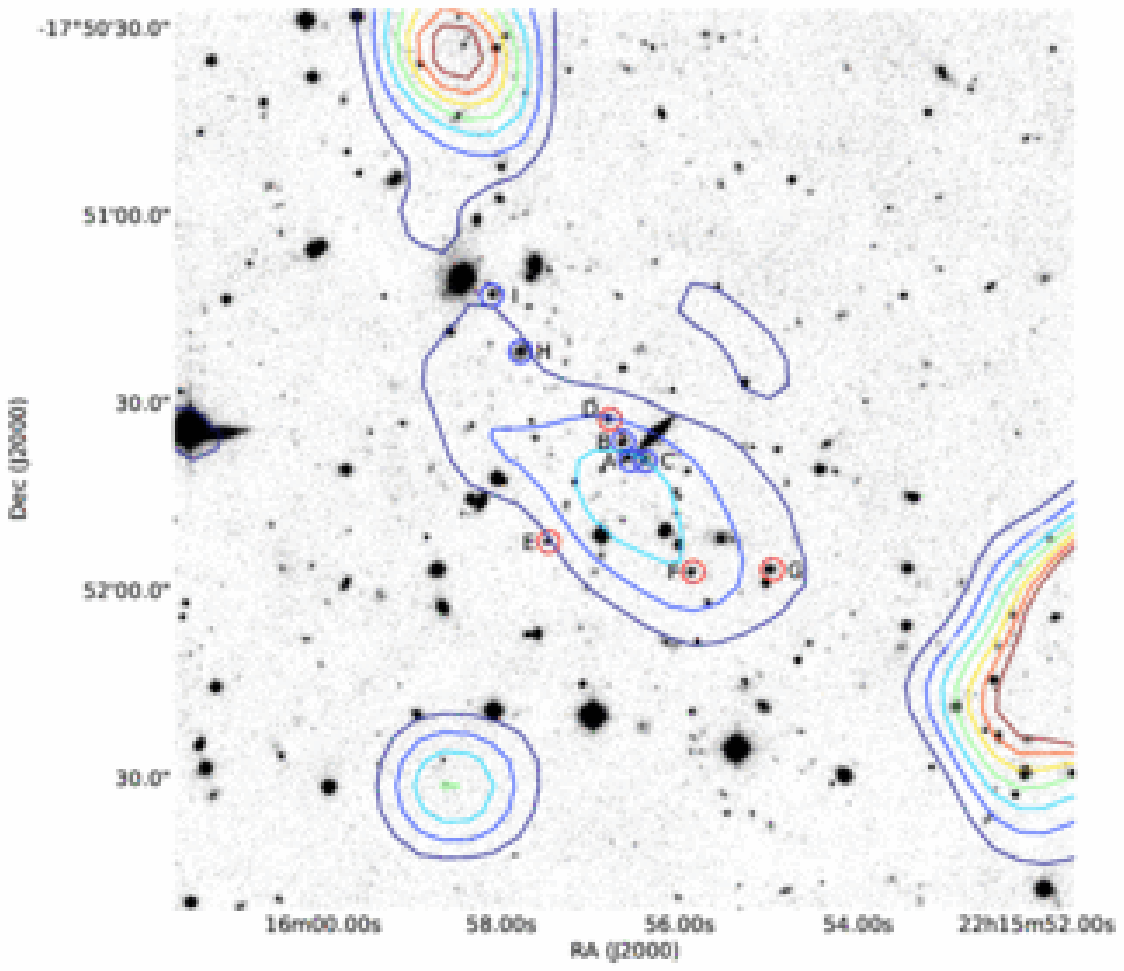}
  \caption{As in Fig. \ref{fig:sky:ext1} for source ext\#5. The associated
  spectra are seen in Fig. \ref{fig:spec:ext5}.}
  \label{fig:sky:ext5}
\end{figure}

\begin{figure}[htb] 
  \begin{center}
   \includegraphics[trim=1cm 0cm 1cm 0cm, clip=true, width=.5\textwidth]{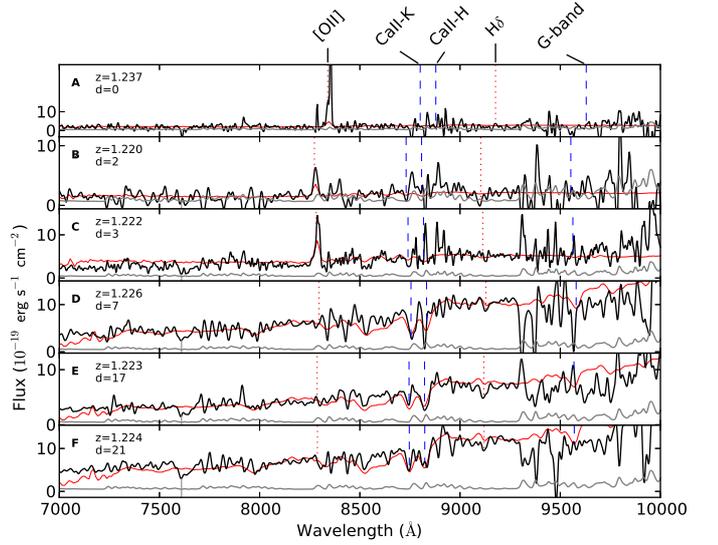} 
  \end{center}
  \caption{Selected spectra for cluster ext\#5 in the observers' frame. Colours
  and labels are as in Figure \ref{fig:spec:ext1}. The positions are seen in
  Figure \ref{fig:sky:ext5}.} 
  \label{fig:spec:ext5} 
\end{figure}

The optical imaging shows a conglomeration of both nearby and distant galaxies.
The CFHT $z'$-band image is shown in Figure \ref{fig:sky:ext5} with X-ray
contour and spectroscopically confirmed cluster members. This cluster has
already been hinted at by \cite{bielby}, however, at a photometric redshift of
$z=1.17$.
The colour-magnitude of the cluster is somewhat peculiar as seen in Figure
\ref{fig:cmd:ext5}. The galaxy closest to the X-ray emission peak is blue and
star-forming. Its membership is confirmed both spectroscopically and
photometrically. Although the colour of the red members is as expected, we note
there seems to be a significant ongoing build up of the red sequence.  These
properties are similar to the cluster found at $z=1.56$ by \cite{renedistant}. 
This second farthest cluster in the field has $12$ concordant redshifts with
$\mid z-z_\text{cluster}\mid \lesssim 0.013$. Figure \ref{fig:spec:ext5} shows
a selection of six members. Mask offsets\footnote{Due to accidental pointing
errors.} during observation caused the signal-to-noise to remain very low.
However, due to the existence of single emission lines in the very centre of
the clusters, the redshift could be established as $z=1.227$. A redshift based
only on one emission line remains tentative; however, the morphology of the
galaxy in combination with its extent on the sky does not render it a likely
possibility that we are seeing a background high-$z$ Ly$\alpha$ emitter.

For this source the X-ray spectrum indicates a redshift lower limit of
$z>0.9$. Both the X-ray and galaxy over-density contours show an NW-SE
elongation.

\subsection{XMMU 221503.6-175215 (ext\#6) $z=0.41$}
\label{sec:ext6}

\begin{figure}[tb]
  \centering
  \includegraphics[trim=.3cm 0cm 1cm 0cm,clip=true,width=.5\textwidth]{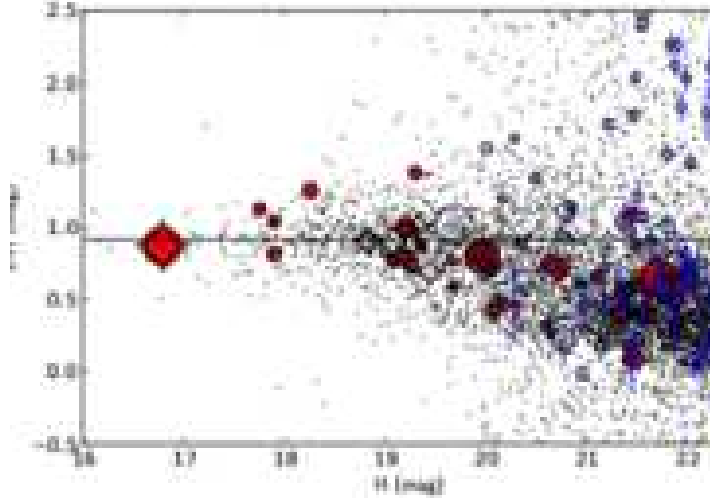}
  \caption{As in Figure \ref{fig:cmd:ext1} for source ext\#6 at $z=0.41$.}
  \label{fig:cmd:ext6}
\end{figure}

\begin{figure}[tbh] 
  \centering 
  \includegraphics[width=.5\textwidth, trim=0.5cm 1cm 2.5cm 2cm, clip=true]{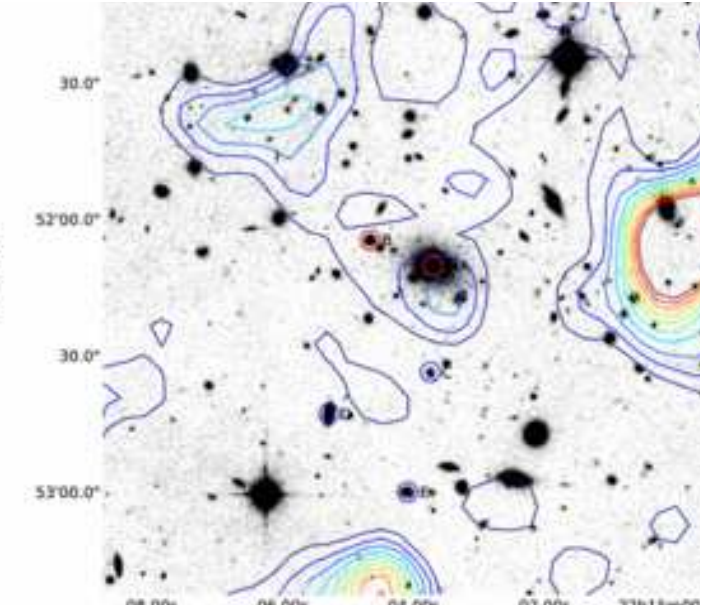}
  \caption{As in Fig. \ref{fig:sky:ext1} for source ext\#6. The associated
  spectra are found in Fig. \ref{fig:spec:ext6}.}
  \label{fig:sky:ext6}
\end{figure}

\begin{figure}[tbh]
  \centering
	\includegraphics[trim=1cm 0cm 1cm 0cm, clip=true, width=.5\textwidth]{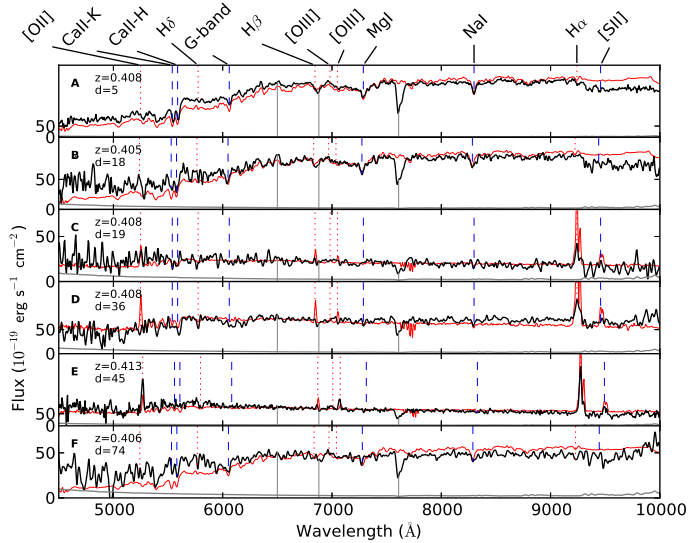}
  \caption{Selected spectra for cluster ext\#6 in the observers' frame. Colours
  are as in Figure \ref{fig:spec:ext1}. Labels refer to identified objects in
  Figure \ref{fig:sky:ext6}.}
  \label{fig:spec:ext6}
\end{figure}

The cluster with the weakest X-ray emission at a redshift of $z=0.410$ can
easily be understood as a sign of large-scale structure. Cluster ext\#6
possibly is a knot in the filaments around the more luminous, hence more
massive structure ext \#1. This work provides the first reference to this
cluster.
The colour-magnitude diagram in Figure \ref{fig:cmd:ext6} shows a well-evolved
red sequence, which both confirms the maturity of the system and rejects the
possibility of a chance-alignment. Since the colour is virtually identical to
that of cluster ext\#1, we have established further proof for the common
redshift of the two clusters. 
We find $14$ spectroscopic members (see Figure \ref{fig:sky:ext6}) for the
clusters of which six are plotted in Figure \ref{fig:spec:ext6}.

The claim to perceive cluster ext\#6 (Figure \ref{fig:zmaps}) as part of the
larger structure around system ext\#1 is supported by the fact that we see
emission line galaxies close to the centre of the cluster. Three of the five
galaxies nearest to the BCG ($<1'$) show at least H$\alpha$ in emission.
Evolved clusters are not expected to show strong signs of star formation, yet
in the outskirts of the largest dark matter potentials, such as cluster ext\#6,
we do expect galaxies to be more active.

\subsection{XMMU 221546.2-174002 (ext\#7) and XMMU 221551.7-173918 (ext\#8)}

\begin{figure}[tb]
  \centering
  \includegraphics[trim=.3cm 0cm 1cm 0cm,clip=true,width=.47\textwidth]{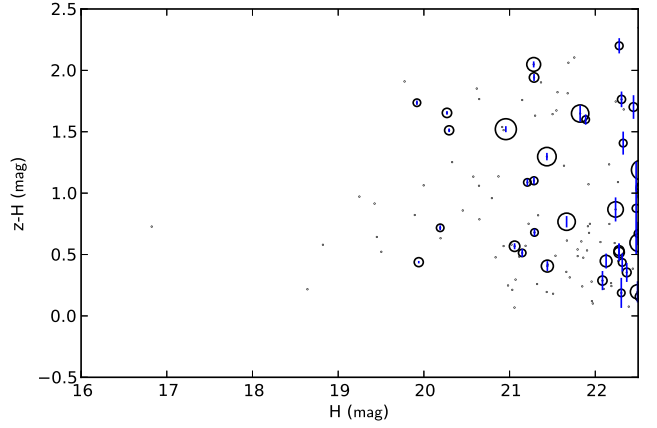}
  \caption{Colour magnitude diagrams around the two
  extended X-ray sources (see also Fig. \ref{fig:cmd:ext8}) showing no galaxy over-density or obvious red
  sequence. Open circles denote all galaxies with good photometry within
  1 arcmin from the X-ray centroid, scaled inversely with their distance to
  this centroid. Likewise, the dots represent objects within 2 arcmin.}
  \label{fig:cmd:ext7}
\end{figure}

\begin{figure}[tb]
  \centering
  \includegraphics[trim=.3cm 0cm 1cm 0cm,clip=true,width=.47\textwidth]{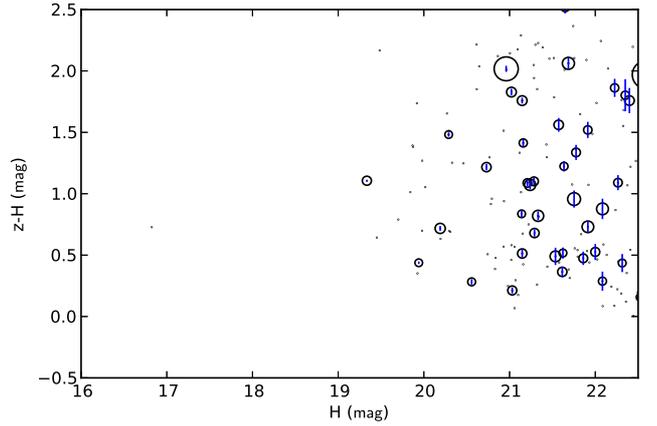}
  \caption{See caption Fig. \ref{fig:cmd:ext7}}
  \label{fig:cmd:ext8}
\end{figure}

\begin{figure}[tb]
  \centering
  \includegraphics[width=.45\textwidth, trim=.5cm 1cm 2.5cm 2cm, clip=true]{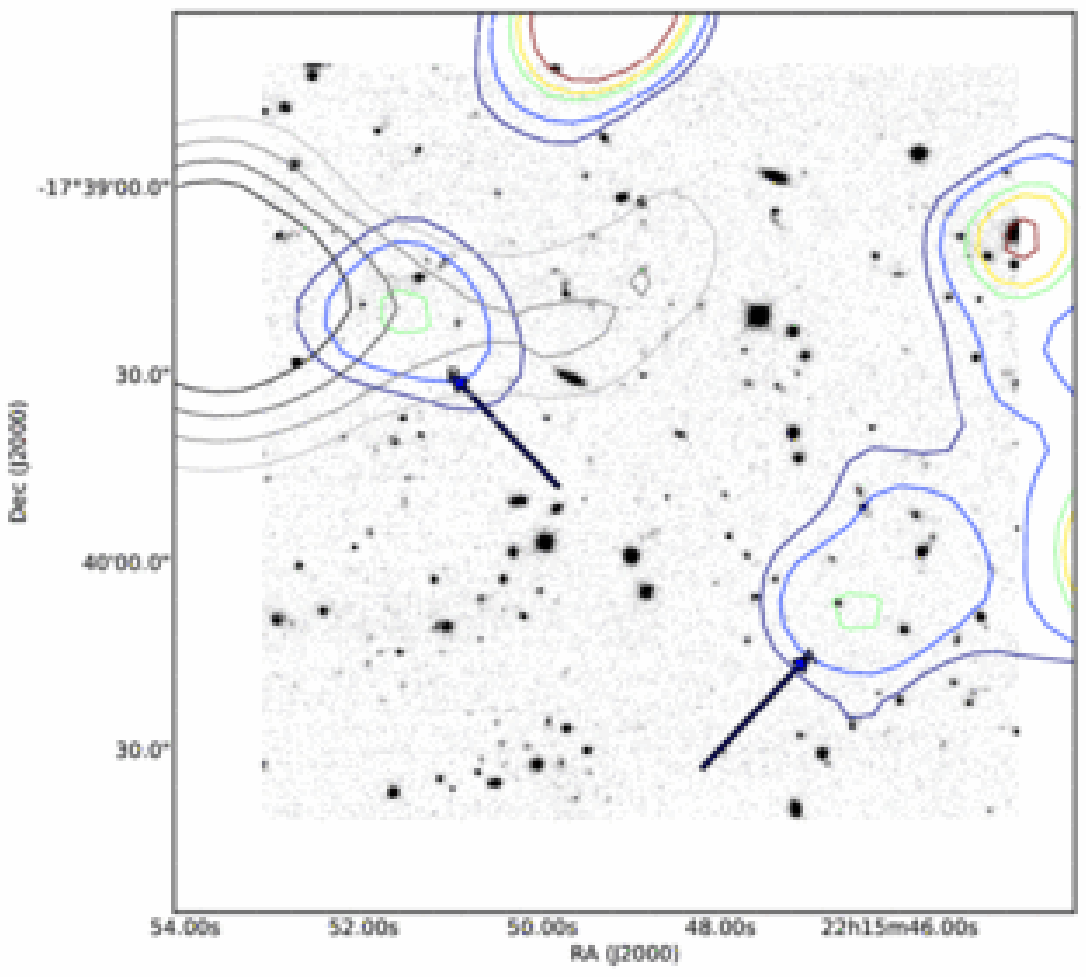}
  \caption{Source ext\#7 and source ext\#8. Stack of NIR $HJK$ bands to
  enhance depth for the two unknown structures. Source ext\#8 is NW and source
  ext\#7 SE of the image centre as pointed out by the arrows. The grey contours
  display the NVSS radio source. Dimension and orientation is as in Fig.
  \ref{fig:sky:ext1}.}
  \label{fig:sky:ext78}
\end{figure}

No clear cluster identification for these X-ray sources can be found in the
optical data. Their vicinity on the sky makes it is natural to discuss these
structures in symbiosis. Remarkably, neither in the very deep CFHT imaging nor
in the WIRCam NIR data do we find a plausible visible counterpart for these
sources as seen in Fig. \ref{fig:sky:ext78}, which rules out the possibility
of a very distant cluster of galaxies within any reasonable redshift limit
(i.e. $z\lesssim 2$). The colour-magnitude diagrams as displayed in Figures
\ref{fig:cmd:ext7} and \ref{fig:cmd:ext8} do not show any hint of a
red sequence. 

Source ext\#7 has the lowest value for EXT\_LIKE (extent likelihood in Table
\ref{extended}) in this sample and could be a blend of faint point sources
and/or associated with source ext\#8. The NVSS radio catalogue \citep{condon}
contains a faint radio source (3.3 mJy at 1.4 GHz) at $18$ arcsecond distance
of source ext\#8. The projection of the two object on the sky (NW versus SE
with respect to the centre of Figure \ref{fig:sky:ext78}) raises the idea of an
associated bipolar phenomenon. 

Furthermore, that both sources were not detected in the cluster's typical
energy band ($0.35-2.4$ keV, see Section \ref{sec:srcdetect}) suggests they may
possess a relatively hard spectrum, which in turn favours another mechanism at
work rather than bremsstrahlung.

Based on the extended nature of the sources, the radio detection, the hardness
of the spectrum, as well as the absence of a clear optical counterpart(s),
little alternative is left but to identify source ext\#8 (and source ext\#7) as
a Compton ghost \citep{fabian}.

\subsection{XMMU 221624.3-173321 (ext\#9)}
\label{sec:ext9}

This source is excluded from further (cosmological) analysis, both for its
being at a large off-axis angle, namely $16$ arcmin (i.e. covered by the \epic
PN camera only), and its likely being a point source.  Nonetheless, at about
$1$ arcmin from source ext\#9 there is a complex X-ray structure (at an
off-axis angle of $16$ arcmin as well), which was not flagged as extended by
our source detection. The two sources are displayed in Figure
\ref{fig:sky:ext9}, centred on the second source, while source ext\#9 is SW
towards the bottom. The structure at R.A. 22:16:20.0 Decl. -17:32:25 has also
been found by our photometric over-density search. It was proposed that it is
at $z\sim0.7$ in the MF sample of \cite{olsen} and in the weak lensing sample
of \cite{bolox}. 

\begin{figure}[tb]
  \centering
  \includegraphics[trim=.3cm 0cm 1cm 0cm,clip=true,width=.5\textwidth]{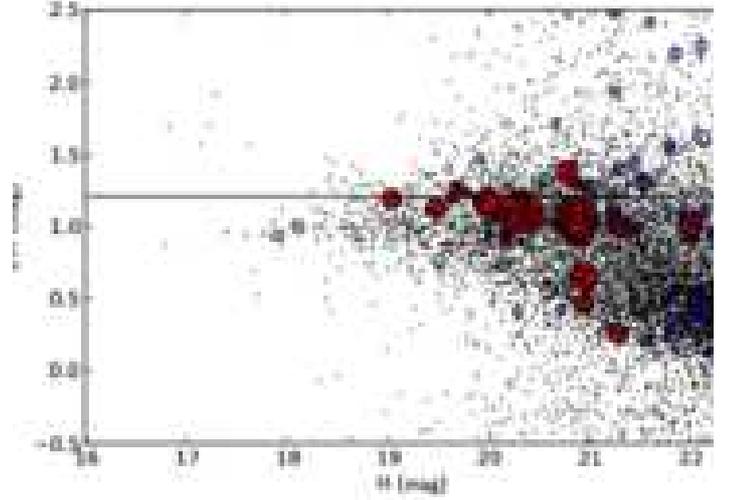}
  \caption{As in Figure \ref{fig:cmd:ext1} for centre coordinates as described
  in Section \ref{sec:ext9} close to source ext\#9. The difficulty to pin-point
  the cluster centre results in an only vague red sequence.}
  \label{fig:cmd:ext9}
\end{figure}

The precise coordinates of the cluster centre (hence the BCG) are problematic
to constrain. The position of the X-ray emission peak is not known exactly due
to contaminating sources and because it is at the very \epic camera's chip
edge. Furthermore, all optical over-density methods have an arcmin-like spatial
resolution (see dark pink contours in Figure \ref{fig:zmaps}).  Therefore, a
tentative BCG is selected at $z=0.71\pm0.07$, with the need to be confirmed by
spectroscopy in the future. The tentative colour-magnitude diagram in Figure
\ref{fig:cmd:ext9} is consistent with a redshift of $z=0.7$.

\begin{figure}[tb]
  \includegraphics[width=.5\textwidth, trim=0.5cm 1cm 2.5cm 2cm, clip=true]{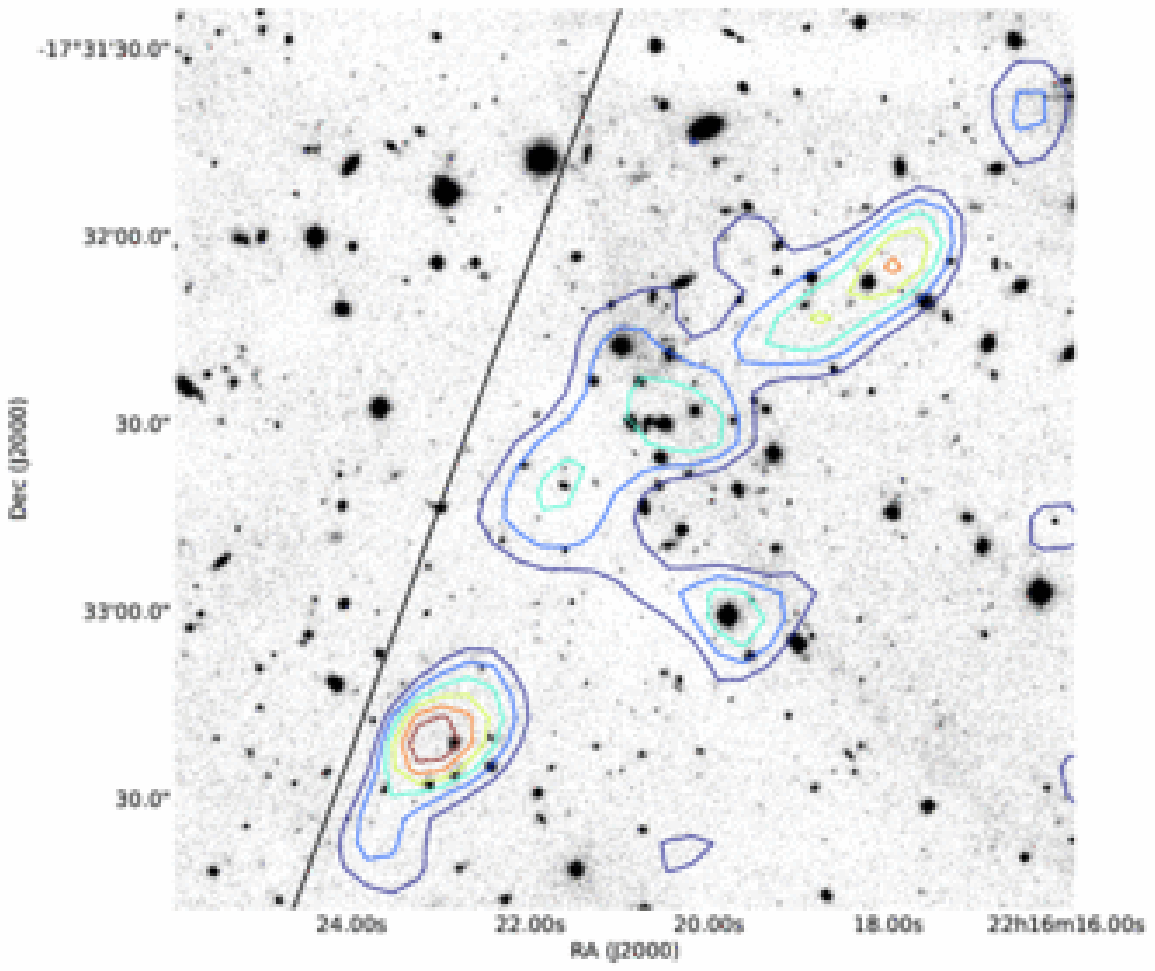}
  \caption{As in Fig. \ref{fig:sky:ext1}, however, the image is centred on the
  cluster of galaxies found close to source ext\#9 (see Section
  \ref{sec:ext9}). The extended source from our source detection is SE of the
  image centre. The black line indicates the end of the \xmm \epic chip.}
  \label{fig:sky:ext9}
\end{figure}

\section{Cluster counts}
\label{sec:cosmo}

\subsection{X-ray sensitivity function}

The advantage of using only X-rays to find clusters of galaxies is the grasp
one has on the selection function. In a rather straightforward manner the X-ray
behaviour of a cluster sample can be mimicked and convolved with a detector
response, thereby obtaining the detection limits for various physical
parameters. In this study, the survey completeness is determined using a
sensitivity function, which is generated by means of Monte-Carlo simulations.
The details of the simulations are presented in \cite{martin} and discussed in
\cite{renexdcp}. An abbreviated description is found in this section.

The probability of detecting extended X-ray sources at a (1) certain flux limit
and (2) a given distance from the optical axis, is computed for ten simulated
beta models ($\beta=\tfrac{2}{3}$) inserted randomly on concentric rings in the
stacked \xmm pointings. During insertion, regions that already contain an
extended source are avoided. If this were not done, we would introduce an
unphysical overlap between clusters at this faint end of the flux scale, which
will result in an over-estimation of the source density when correcting for
this unrealistic projection effect.  The simulated clusters cover a range of
$25$ bins in both core radii and photon counts. This results in $3\,125$
simulations, since the procedure is repeated five times. The core radii
distribution assumed is taken from \cite{vik98}. The cluster sample in
\cite{vik98} consists of 200 clusters observed with ROSAT with redshift
information. The redshifts range from $0.015-0.73$ with $<z>=0.28$. The
advantage of this particular sample is that the core radius is computed by
applying $\beta=0.67$ as in our analysis. The core radii derived are scaled to
how they would appear at $z=1$ using the geometrical angular distance
dependency on redshift. No further evolution is assumed. The mean core radius
at $z=1$ is $\sim14$ arcsec.

\begin{figure}[thb] 
  \centering
  \includegraphics[trim=0.5cm 0cm 1.5cm 1cm, clip=true, width=.5\textwidth]{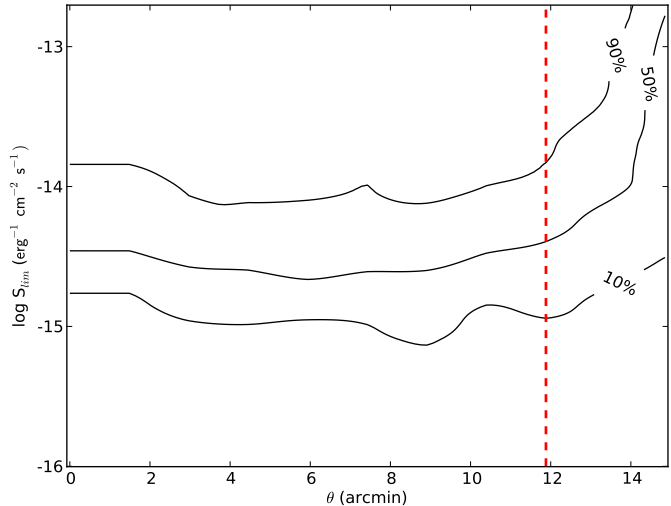}
  \caption{Survey completeness as a function of off-axis
  angle and limiting flux. The 10\%, 50\% and 90\% levels are indicated. The
  dashed red line indicates the restriction put on the off-axis angle.}
  \label{fig:sim:a}
\end{figure}

\begin{figure}[bth] 
  \includegraphics[trim=0.5cm 0cm 1.5cm 1cm, clip=true, width=.5\textwidth]{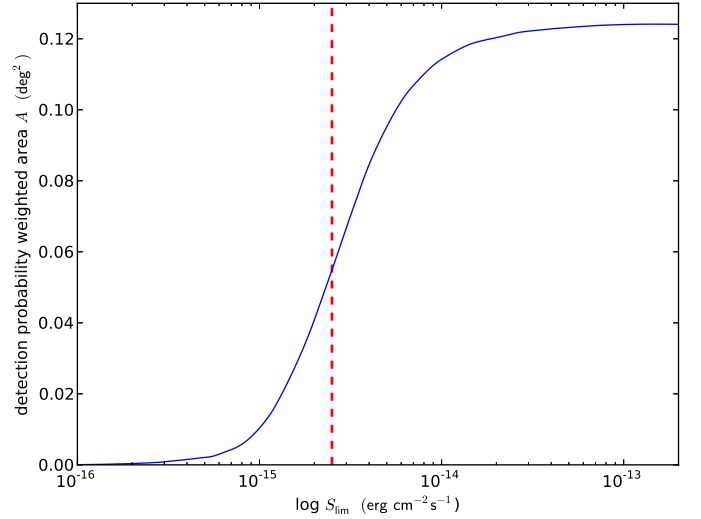}
  \caption{Cumulatively sensitivity-weighted surface areas as a
  function of limiting flux, integrated over all off-axis angles. The
  short-dashed red line indicates the deduced limiting flux of
  $2.5\cdot10^{-15}\;\text{erg cm}^{-2}\,\text{s}^{-1}$ at which $50\%$
  completeness is reached as indicated in Figure \ref{fig:sim:a}.}
  \label{fig:sim:b}
\end{figure}

Source detection is performed for every simulation with {\tt \sas\ v.6.5} in
the multiple energy bands simultaneously with the same parameters as mentioned
in Section \ref{sec:srcdetect}. The resulting image displaying the survey
completeness level as a function of limiting flux ($S_\text{lim}$) and off-axis
angle ($\theta$) is shown in Figure \ref{fig:sim:a}. We notice that at the
indicated completeness levels, the curve is rather flat between ca. $2-12$
arcmin and in particular the $50\%$ curve, which we take as the completeness
level for our survey. Towards the centre of the pointing, however, an extended
structure (cluster ext\#1) affects the sensitivity in this observation, while
on the outskirts the vignetting and the very end of the chip catastrophically
affect the detectability of clusters. At the $50\%$ level, corresponding to a
flux of $2.5\cdot10^{-15}\;\text{erg cm}^{-2}\,\text{s}^{-1}$, we are complete.
All clusters detected have (source detection) fluxes above this level. For our
$\log N-\log S$ analysis, none of the objects ext\#1-\#6 needs to be excluded:

\begin{equation}
  A(S_\text{lim})=\sum_{i=1}^{12} \sum_{S_\text{lim}} A(\theta_i)\cdot P(\theta_i,S_\text{lim}).
  \label{eq:eff_area}
\end{equation}

Dividing the \lbqs\ observation into rings, we can fold the geometric area with
the survey sensitivity at each off-axis angle $\theta$. We limit the radius to
$\leq 12$ arcmin. Beyond this radius the sensitivity drops significantly as
seen in Figure \ref{fig:sim:a}. This area restriction also ensures the coverage
by all three \epic cameras. One extended source (ext\#9) is excluded as a
result of this angle limit. The cumulative effective surface area, or detection
probability weighted area $A(S_\text{lim})$, computed following equation
\eqref{eq:eff_area} is displayed in Figure \ref{fig:sim:b}. Owing to our limit
on the maximum radius, we obtain an survey area of $\sim 0.13\,\Box^\circ$.
$S_\text{lim}$ denotes the limiting flux. This is the correction curve to be
convolved with the clusters' number counts in the purely geometrical survey
area, resulting in the survey-independent cosmological cluster count in flux
bins.

\subsection{ \textrm{log}N--\textrm{ log}S}
\label{sec:lnls}

\begin{figure}[tbp] 
  \centering
  \includegraphics[trim=0cm 0cm 0cm 0cm, clip=true, width=.5\textwidth]{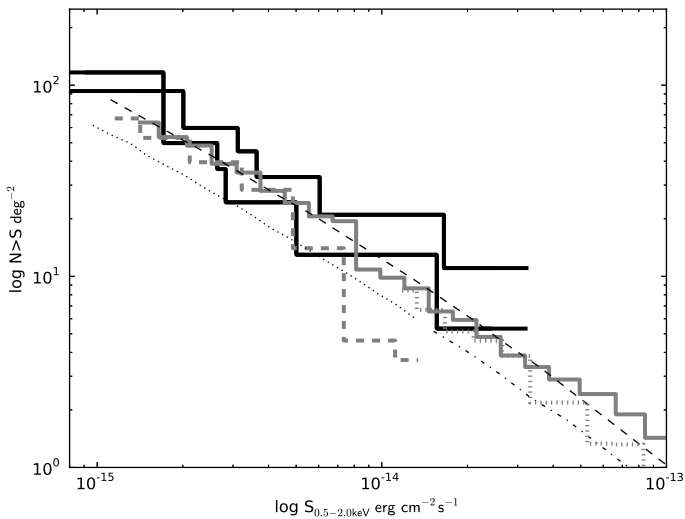}
  \caption{Cosmological $\log N-\log S$ for clusters of galaxies. The black
  discrete curve shows the confidence interval resulting from measured source
  detection flux errors and an additional $\sqrt{N}$ on the number count from
  this work.  A comparison to comparable surveys is made. The curves show the
  model from \cite{piero02} for no evolution (dashed). The dot-dashed and
  dotted lines show the measured ROSAT brightest cluster sample
  \citep[BCS,][]{ebeling98} clusters and their extrapolation, respectively. The
  discrete lines display the findings from COSMOS (solid) \citep{cosmos}, SXDF
  (dashed) \citep{sxdf}, and XMM-BCS (dotted) \citep{suhada12}.}
  \label{fig:detect} 
\end{figure}

The computed sensitivity function is used to predict the number of X-ray
luminous clusters in a certain survey area. We convolve the probability of
detecting clusters in concentric rings around the focal axis with the enclosed
area (Figure \ref{fig:sim:b}). The observed flux distribution is normalised
to the probability weighted survey area to obtain the expected number of
galaxy clusters per flux bin. The $\log N-\log S$ graph is presented in Figure
\ref{fig:detect}. The empirical values are compared to the expected numbers
simulated by \cite{piero02,rdcs}, which in turn proved to be a good
approximation for the extended X-ray source distribution in the COSMOS
\citep{cosmos} and the Subaru-\xmm deep field \citep[SXDF,][]{sxdf}.

When making the link between the derived flux limits and our measured fluxes,
we have to bear in mind that the simulated clusters cover a range of core
radii as they appear at $z=1$. Therefore we compute $S_\text{500}$ for
hypothetical clusters at $z=1$ with $r_\text{c}=14$ arcsec, using the iterative
approach, which spans the flux range from our simulations. The mean conversion
factor between $S_\text{lim}$ and $S_\text{500}$ is 0.8, which we apply to our
limiting fluxes.

The intrinsic errors resulting from the method used to obtain $S_\text{500}$
do not enter into Figure \ref{fig:detect}, but only the spectral flux
errors.  Furthermore, since the distribution of clusters on the sky is
Poissonian, we add an additional error of $\sqrt{N}$ on the number counts.

The number counts in flux bins from this study are, among others and within
errors, compatible with the findings from the SXDF and COSMOS. Likewise, using
spectral or source detection fluxes has no dramatic effect on the results. For
low-flux objects, the scenario in which there is {\it no evolution} in the XLF
for extended sources \citep{piero02} is favoured.

\section{Summary}
\label{sec:sum}

This work presents the detection and study of extended X-ray sources in a deep
(250 ksec) \xmm field covering a field of view of $\sim 14$ arcmin across.
The nature of the sources has been identified. We have valiantly achieved
``five at one blow'', since this single field has allowed us to discover
five new spectroscopically confirmed clusters. In the following we summarise
the most essential findings from this study.

\begin{itemize}
  \item We performed a robust source detection on deep, stacked \xmm
    observations targetting \lbqs, revealing nine extended sources with
    conservative detection thresholds. Six of these sources prove to be
    clusters of galaxies, five of which are new spectroscopically confirmed
    discoveries from this work alone. The sixth detected cluster was already
    known from \cite{stanford}. The first unconfirmed extended source, at $>14$
    arcmin off-axis, is a confused point source. The remaining two sources
    could not be identified. Since the detection of these sources is dominated
    by the hard energy bands, this suggests that another mechanism than
    bremsstrahlung is responsible for the X-ray emission from these two
    sources, which possibly leads at least one of them to be a Compton ghost.
  \item A complementary qualitative optical/IR completeness study, allowed by
    very deep CFHTLS data (i.e.  $z'_\text{lim}=26.0\,H_\text{lim}=24.6$), has
    confirmed that all X-ray luminous clusters to be associated with an optical
    over-density, in agreement with the spectroscopic redshift. One new,
    significant over-density could not be correlated with any X-ray emission,
    hinting at a projection effect in the optical search method.
  \item The X-ray selection function has been computed through extensive
    Monte-Carlo simulations.  The approach of retrieving generated $\beta$
    models inserted into the observations has given us a $50\%$ completeness
    flux limit of $\sim 2.5\cdot10^{-15}\,\text{erg s}^{-1}$ over most of the
    field of view. We constructed a $\log N-\log S$ based on a complete sample
    of spectroscopically confirmed sources down to weak fluxes in a precisely
    determined survey area with a known selection function. Our result agrees
    with the findings from the COSMOS and the SXDF.  Comparing the number
    counts with the models from \cite{piero02}, our data seems to favour no
    evolution in the XLF in the faint flux range we covered.
  \item By spectroscopic follow-up of some X-ray point sources, we
    serendipitously discovered three new AGN. Two of these objects reside at
    the concordant redshift of the cluster \citep{stanford} at $z=1.45$. The
    other quasar has a redshift of $z=4.20$, so is among the handful of most
    distant X-ray-selected AGN known today \citep{fiore12}.
\end{itemize}

In this deep field study, good photometric data have allowed us to
scan the sky in multiple wavelengths to study the galaxy counterparts of
extended X-ray emission. We emphasise that photometric redshifts
with errors $\Delta z < 0.1$ are still too expensive to date for routinely
performing this kind of analysis. With respect to the upcoming X-ray mission
eROSITA \citep{erosita}, future optical surveys such as LSST
\citep{lsst}, the ESO/Vista project VHS \citep{vista}\footnote{data sharing
agreement with DES (Dark Energy Survey)}, and PanStarrs \citep{panstarrs}
will obtain sufficient depth and cover enough area to identify a wide range of
distant and faint X-ray clusters. This work, therefore, could very well be
extended to a much larger area of the sky within the next decade.

In the near future, as the photometric and spectroscopic follow-up campaigns
for the XDCP is becoming complete, we are already compiling catalogues of
distant clusters to extend the present work.

\begin{acknowledgements}
  AdH was supported by the DFG under fund no. Schw536/24-2.  GL acknowledges
  support by the Deutsches Zentrum f\"ur Luft- und Raumfahrt (DLR) under
  contract no.~FKZ 50 OX 0201 and 50 QR 0802. \\
  JPD is supported by by the German Science Foundation (DFG) through the
  Transregio 33 and the Cluster of Excellence ``Origin and Structure of the
  Universe'', funded by the Excellence Initiative of the Federal Government of
  Germany, EXC project number 153.\\
  DP acknowledges the kind hospitality at the MPE.\\
  This research made use of (1) APLpy, an open-source plotting package for
  Python hosted at http://aplpy.github.com and (2) IPython, an interactive
  python shell \citep{ipython}. \\
  We are grateful to the CFHT survey team for conducting the observations and
  the TERAPIX team for developing software used in this study. We acknowledge
  use of the Canadian Astronomy Data Centre, which is operated by the Dominion
  Astrophysical Observatory for the National Research Council of Canada's
  Herzberg Institute of Astrophysics. It is based on observations obtained with
  WIRCam, a joint project of CFHT,Taiwan, Korea, Canada, France, at the
  Canada-France-Hawaii Telescope (CFHT) which is operated by the National
  Research Council (NRC) of Canada, the Institute National des Sciences de
  l'Univers of the Centre National de la Recherche Scientifique of France, and
  the University of Hawaii. This work is based in part on data products
  produced at TERAPIX, the WIRDS (WIRcam Deep Survey) consortium, and the
  Canadian Astronomy Data Centre. This research was supported by a grant from
  the Agence Nationale de la Recherche ANR-07-BLAN-0228. This research has
  made use of the NASA/IPAC Extragalactic Database (NED), which is operated by
  the Jet Propulsion Laboratory, California Institute of Technology, under
  contract with the National Aeronautics and Space Administration.\\ 
  M. L.  thanks the European Community for the Marie Curie research training
  network ``DUEL'' doctoral fellowship MRTN-CT-2006-036133.\\

\end{acknowledgements}


\Online
\begin{appendix}

\onecolumn

%
%

\end{appendix}
\end{document}